\documentclass[twocolumn]{aastex631}

\usepackage{amsmath}

\definecolor{fuchsia}{rgb}{0.54, 0.17, 0.89}
\definecolor{azure}{rgb}{0.0, 0.5, 1.0}
\definecolor{pgreen}{rgb}{0.12, 0.3, 0.17}
\definecolor{alizarin}{rgb}{0.82, 0.1, 0.26}

\newcommand{\kms}{{\rm km~s^{-1}}}
\newcommand{\oii}{[\textrm{O}~\textsc{ii}]}
\newcommand{\oiii}{[\textrm{O}~\textsc{iii}]}
\newcommand{\oiiir}{[\textrm{O}~\textsc{iii}]_{\lambda5007}}

\newcommand{\oiiif}{$[\textrm{O}~\textsc{iii}]_{\lambda4363}$}
\newcommand{\nii}{[\textrm{N}~\textsc{ii}]}

\newcommand{\simgt}{\,\rlap{\lower 3.5 pt \hbox{$\mathchar \sim$}} \raise
1pt \hbox {$>$}\,}
\newcommand{\simlt}{\,\rlap{\lower 3.5 pt \hbox{$\mathchar \sim$}} \raise
1pt \hbox {$<$}\,}
\newcommand{\Msun}{M_{\odot}}

\newcommand{\logm}{\log M_*/\Msun}

\newcommand{\logoh}{$\log {\rm (O/H)}$}

\newcommand{\ha}{${\rm H\alpha}$}
\newcommand{\hb}{${\rm H\beta}$}
\newcommand{\hg}{${\rm H\gamma}$}
\newcommand{\hd}{${\rm H\delta}$}

\newcommand{\Z}{{\rm [Z/H]}}

\newcommand{\nsrcz}{9} 
\newcommand{\nsrc}{28} 
\newcommand{\nsrcext}{16} 
\newcommand{\nsrcj}{6} 
\newcommand{\nsrcn}{10} 
\newcommand{\nsrcall}{25} 

\accepted{May 29, 2024}

\submitjournal{ApJ}

\shorttitle{Metallicity of Galaxies at $3<z<9.5$}
\shortauthors{Morishita, Stiavelli, et al.}

\graphicspath{{./}{figures/}}

\begin{document}

\title{
Diverse Oxygen Abundance in Early Galaxies Unveiled by Auroral Line Analysis with JWST
}

\correspondingauthor{Takahiro Morishita}
\email{takahiro@ipac.caltech.edu}

\author[0000-0002-8512-1404]{Takahiro Morishita}
\affiliation{IPAC, California Institute of Technology, MC 314-6, 1200 E. California Boulevard, Pasadena, CA 91125, USA}

\author[0000-0001-9935-6047]{Massimo Stiavelli}
\affiliation{Space Telescope Science Institute, 3700 San Martin Drive, Baltimore, MD 21218, USA}

\author[0000-0002-5926-7143]{Claudio Grillo}
\affiliation{Dipartimento di Fisica, Università degli Studi di Milano, Via Celoria 16, I-20133 Milano, Italy}
\affiliation{INAF - IASF Milano, via A. Corti 12, I-20133 Milano, Italy}

\author[0000-0002-6813-0632]{Piero Rosati}
\affiliation{INAF - OAS, Osservatorio di Astrofisica e Scienza dello Spazio di Bologna, via Gobetti 93/3, I-40129 Bologna, Italy}
\affiliation{Dipartimento di Fisica e Scienze della Terra, Università degli Studi di Ferrara, Via Saragat 1, I-44122 Ferrara, Italy}

\author[0000-0003-2497-6334]{Stefan Schuldt}
\affiliation{Dipartimento di Fisica, Università degli Studi di Milano, Via Celoria 16, I-20133 Milano, Italy}
\affiliation{INAF - IASF Milano, via A. Corti 12, I-20133 Milano, Italy}

 \author[0000-0001-9391-305X]{Michele Trenti}
 \affiliation{School of Physics, University of Melbourne, Parkville 3010, VIC, Australia}
 \affiliation{ARC Centre of Excellence for All Sky Astrophysics in 3 Dimensions (ASTRO 3D), Australia}

\author[0000-0003-1383-9414]{Pietro~Bergamini}
\affiliation{Dipartimento di Fisica, Università degli Studi di Milano, Via Celoria 16, I-20133 Milano, Italy}
\affiliation{INAF - OAS, Osservatorio di Astrofisica e Scienza dello Spazio di Bologna, via Gobetti 93/3, I-40129 Bologna, Italy}

\author[0000-0003-4109-304X]{Kit Boyett}
\affiliation{School of Physics, University of Melbourne, Parkville 3010, VIC, Australia}
\affiliation{ARC Centre of Excellence for All Sky Astrophysics in 3 Dimensions (ASTRO 3D), Australia}

\author[0000-0001-7583-0621]{Ranga-Ram Chary}
\affiliation{IPAC, California Institute of Technology, MC 314-6, 1200 E. California Boulevard, Pasadena, CA 91125, USA}

\author[0000-0003-4570-3159]{Nicha Leethochawalit}
\affiliation{National Astronomical Research Institute of Thailand (NARIT), Mae Rim, Chiang Mai, 50180, Thailand}

\author[0000-0002-4140-1367]{Guido Roberts-Borsani}
\affiliation{Department of Astronomy, University of Geneva, Chemin Pegasi 51, 1290 Versoix, Switzerland}

\author[0000-0002-8460-0390]{Tommaso Treu}
\affiliation{Department of Physics and Astronomy, University of California, Los Angeles, 430 Portola Plaza, Los Angeles, CA 90095, USA}

\author[0000-0002-5057-135X]{Eros~Vanzella}
\affiliation{INAF -- OAS, Osservatorio di Astrofisica e Scienza dello Spazio di Bologna, via Gobetti 93/3, I-40129 Bologna, Italy}



\begin{abstract}
We present deep JWST NIRSpec observations in the sightline of MACS~J1149.5+2223, a massive cluster of galaxies at $z=0.54$\,. We report the spectroscopic redshift of \nsrc\ sources at $3<z<9.1$, including \nsrcz\ sources with the detection of the \oiiif\ auroral line. Combining these with \nsrcext\ \oiiif-detected sources from publicly available JWST data, our sample consists of \nsrcall\ galaxies with robust gas-phase metallicity measurements via the direct method. We observe a positive correlation between stellar mass and metallicity, with a $\sim0.5$\,dex offset down below the local relation. Interestingly, we find a larger than expected scatter of $\sim0.3$\,dex around the relation, which cannot be explained by redshift evolution among our sample or other third parameters. 
The scatter increases at higher redshift, and we {tentatively} attribute this to the enrichment process having higher stochasticity due to shallower potential wells, more intense feedback processes, and a higher galaxy merger rate.
Despite reaching to a considerably low-mass regime ($\logm\sim7.3$), our samples have metallicity of \logoh\,$+12\simgt 7$, {\it i.e.} comparable to the  most metal poor galaxies in the local Universe. The search of primordial galaxies may be accomplished by extending toward a lower mass and/or by investigating inhomogeneities at smaller spatial scales.
Lastly, we investigate potential systematics caused by the limitation of JWST's MSA observations. Caution is warranted when the target exceeds the slit size, as this situation could allow an overestimation of ``global" metallicity, especially under the presence of strong negative metallicity gradient.
\end{abstract}



\section{Introduction} \label{sec:intro}

The gas cycle within a galaxy system carries an important role. The ejecta from previously formed stars enrich surrounding gas, some fraction of which is subject to be used in the following star formation cycles, leading to further enhancement in chemical abundances. The accumulation of stars formed in such a cyclical way represents a major component of galaxies in a later epoch, culminating with an established relation of metal contents and other fundamental galaxy properties, such as stellar mass, seen in the local universe \citep[e.g.,][]{tremonti04,gallazzi05,andrews13,kirby13,curti20}. Observations have reported the presence of the relationship out to $z\sim3$, but with an offset to the local relation \citep[e.g.,][]{maiolino08,zahid11,henry13,steidel16,sanders21}.

However, galaxies are by no means closed systems, and thus chemical enrichment on a galaxy lifetime scale is multifaceted. Star formation, ejection of enriched gas by stellar feedback and accretion of pristine gas from cosmic web have a complex interplay that ultimately defines a unique chemical enrichment history for a given object. 
The situation is expected to be even more complicated at earlier times, where the gas infall rate is higher, star formation is more intense, and the gravitational potential well is not deep owing to the lower mass of (proto)galaxies. As such, extending the observational frontier toward higher redshift is critical to our understanding of the origin of such a tight correlation between metallicity, which is often represented by oxygen abundance measurements, and other galaxy properties.

With the advent of JWST \citep{gardner23}, galaxy interstellar medium studies are now possible at $z\simgt4$ \citep{Curti2023,Heintz2023,Laseter2023,nakajima23,sanders23,stiavelli23}. The unparalleled JWST sensitivity can capture spectral features of high redshift galaxies at rest-frame optical wavelengths, such as those from oxygen and hydrogen, enabling a fair comparison to the variety of well-established lower redshift studies in the literature. 
Most importantly, the O/H abundance can be inferred from the electron temperature ($T_e$), which in turn is measured from the ratio of the 
\oiiif\ auroral emission line to $\oiiir$. This measurement, often referred to as {\it direct} method, enables robust determination of oxygen abundances. In contrast, most of the pre-JWST high-$z$ studies adopt indirect metallicity inference, e.g., $R23$ \citep[][]{pettini04, maiolino08,marino13,curti20}. Such indirect methods are in many cases constructed and calibrated from local measurements, and thus have limitation, e.g., are applicable only within a moderate metallicity range $\gtrsim8$ \citep[e.g.,][]{curti20}. As such, their application to high-$z$ galaxies is potentially subject to systematic biases.

Auroral lines are, however, faint. For example, \oiiif\ is of the order of $\sim1/100$ to \oiii$_{\lambda 5007}$, hence the direct oxygen measurement was previously limited to a smaller fraction of $z>0$ galaxies, and to only a handful at $z\sim2$--3 \citep{sanders16,sanders20,gburek19}. On the other hand, \oiiif\ intensity increases as the electron temperature increases. With galaxies being less enriched and hosting more intense star formation, the application of the direct method becomes relatively more favorable within the higher redshift range enabled by JWST.

In this study, we present our Cycle~1 GTO observations \citep[PID~1199;][]{stiavelli23}, executed along the sightline of MACS~J1149.5+2223, a massive cluster of galaxies at $z=0.54$\,. We report the detection of the \oiiif\ line in \nsrcz\ galaxies, out of $\nsrc$ sources at $3<z<9.1$ spectroscopically confirmed in the program. We combine this new sample with spectroscopic data for an additional 16 galaxies with \oiiif\ available in the literature, aiming to obtain robust oxygen abundance measurements of these galaxies and to investigate their relation with global properties. In Sec.~\ref{sec:data} we present our data reduction, followed by our spectroscopic and photometric analyses in Sec.~\ref{sec:ana}. We then characterize their physical properties and investigate the relationship between stellar mass and oxygen abundance in Sec.~\ref{sec:res}. We discuss the inferred properties of early galaxies in the context of galaxy evolution in Sec.~\ref{sec:disc}. Where relevant, we adopt the AB magnitude system \citep{oke83,fukugita96}, cosmological parameters of $\Omega_m=0.3$, $\Omega_\Lambda=0.7$, $H_0=70\,\kms\, {\rm Mpc}^{-1}$, and the \citet{chabrier03} initial mass function. Throughout the manuscript, we use the term metallicity to refer to oxygen abundance.

\section{Data}\label{sec:data}

\subsection{NIRSpec/MSA Spectroscopic Observations}\label{sec:msa}
The NIRSpec/MSA observations were configured with two medium-resolution gratings (G235M and G395M), aiming to capture rest-frame optical emission lines of galaxies at $z\sim3$--9. Our NIRSpec observations were executed over four different visits. The first two visits, \#20 and \#22, were taken at position angle PA=257.766\,degree with G235M and G395M, respectively. In the second group of visits, the same grating pair was used, but with a slightly different position angle, namely PA=259.660\,degree for visit \#21/G235M and PA=256.766\,degree for visit \#23/G395M. The total science time, excluding overhead, is 16632\,sec for each grating, though not all sources have the full exposure coverage due to the mask design.

We designed a slit mask for each of the four MSA visits. The parent source catalog was constructed by using the existing HST imaging data, including CLASH, HFF, GLASS, and BUFFALO \citep{postman12,treu15,lotz17,kelly18,steinhardt20}. Spectroscopically redshift measurements available in the literature were also incorporated \citep[][Schuldt et al., in prep]{grillo16,treu16,shipley18}. In our mask design, the highest priority was given to JD1, a previously known spec-$z$ source at $z=9.1$ \citep{hashimoto18,hoag19}, followed by $z>6$ galaxies presented in \citet[][]{zheng17} and then $z>1$ galaxies. The sources in our catalog were firstly identified based on the existing WFC3-IR-stack detection. However, since the NIRSpec coverage is larger than the WFC3-IR footprint, we expanded our source coverage to the ACS F606W+F814W footprint, which offers $z\simlt4$ dropout sources as fillers. For each source, the default 3 shutter slitlets were assigned, enabling exposure at three positions separated for $0.\!''46$ from each. We selected the default (Entire Open Shutter Area) for source allocations in APT, which secures to place the source center position within 35\,milliarcsec from the shutter boundary. Including fillers and open shutters in empty sky regions, our MSA masks consisted of in total 341 unique sources.

For the MSA data reduction, we use {\tt msaexp}\footnote{https://github.com/gbrammer/msaexp} following \citet{morishita23b}. The one-dimensional spectrum is extracted via optimal extraction. The source light profile along the cross-dispersion direction is directly measured using the 2-d spectrum. For sources with faint continuum or with any significant contamination (i.e. from a failed open shutter), we visually inspect the 2-d spectrum and manually define the extraction box along the trace where any emission lines are identified.

\begin{figure*}
\centering
    \includegraphics[width=1\textwidth]{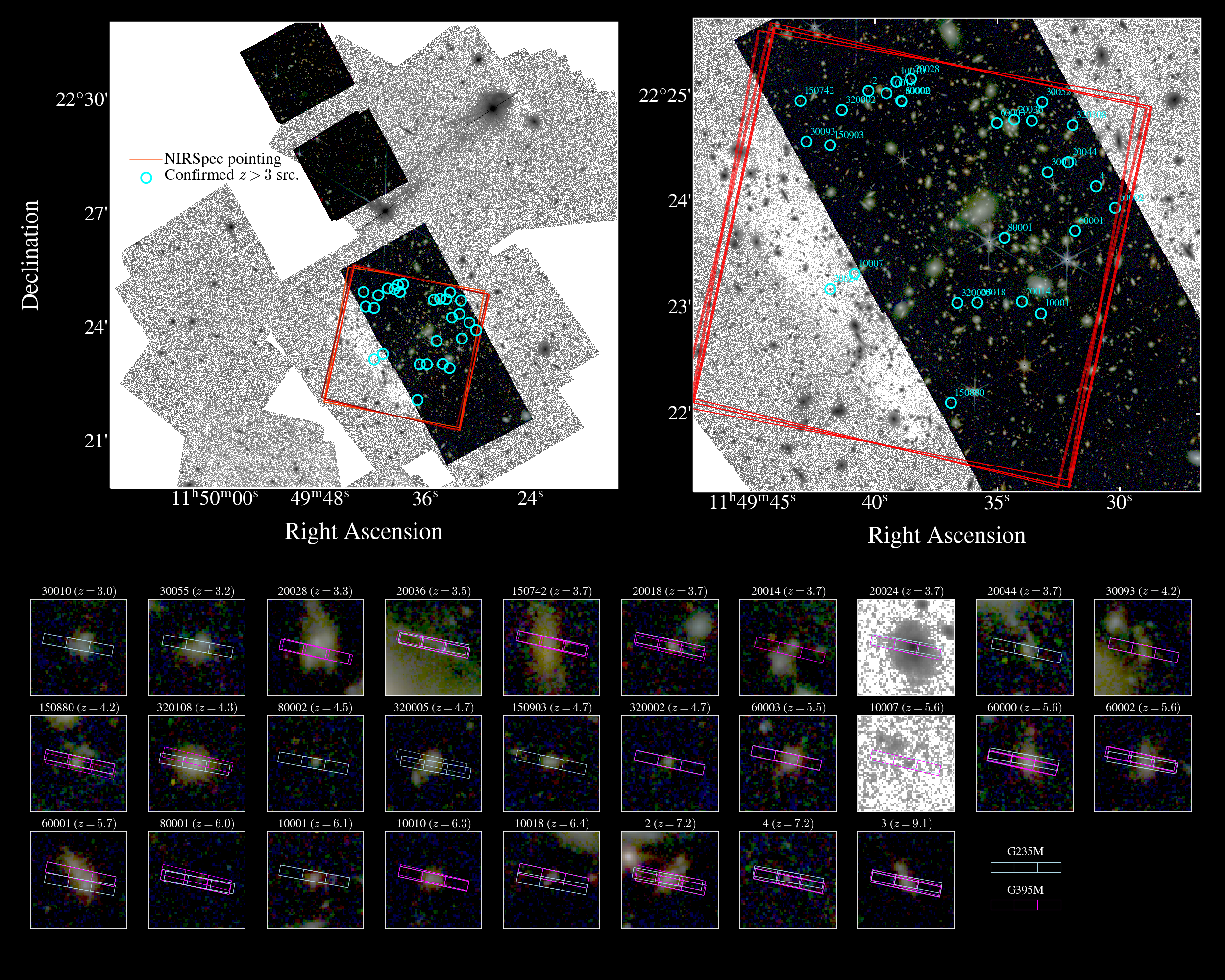}
	\caption{
 (Left):
 Pseudo RGB images of NIRCam in the cluster pointing and the parallel field are shown (F115W+F150W/F200W+F277W/F356W+F444W for blue/green/red). The confirmed $z>3$ sources in our program are marked (cyan). The HST/ACS images are shown in the background. The Hubble Frontier Field's parallel pointing is located to the South of the image (not shown here). NIRSpec footprints are marked (red solid rectangles).
 (Right): Zoomed view around the NIRSpec FoV. 
 (Bottom): Cutout RGB stamps of $z>3$ sources are shown ($2''\!\times2''$). The positions of MSA shutters are overlaid.
 }
\label{fig:mosaic}
\end{figure*}

\subsection{Imaging Data Reduction and Photometry}\label{sec:data-jwst}
NIRCam imaging observations were executed on June 6 -- 8 2023 as part of the same program (Visit \#5), with six filters configured (F115W, F150W, F200W, F277W, F356W, and F444W) with the {\tt MEDIUM8} readout mode, totaling in $\sim1.1$\,hr on-source exposure in each filter. 

We follow the same steps for JWST image reduction presented in \citet{morishita23}. Briefly, we retrieve the raw-level images from the MAST archive and reduce those with the official JWST pipeline {(ver.1.10.0, with {\tt jwst\_1100.pmap})}, with several customized steps including $1/f$-noise subtraction and snowball masking. The final drizzled images are aligned with the same WCS coordinate as for the HST images, with the pixel scale set to $0.\!''03$. In Fig.~\ref{fig:mosaic}, we show the NIRCam and NIRSpec coverage, on top of the footprint of the existing HST images.

We identify sources in the detection image using SExtractor \citep{bertin96}, and measure their fluxes on the psf-matched mosaic images with a fixed aperture of radius $r=0.\!''16$\,. The measured aperture-based fluxes of each source is then scaled by a single factor, defined as ${\rm flux_{auto, F444W}/flux_{aper, F444W}}$, to the total flux. With this approach, colors remain as those measured in the aperture, whereas the fluxes are scaled to the total quantities for subsequent analysis (i.e. derivation of stellar mass and star formation rate).

\section{Analysis}\label{sec:ana}

\begin{figure*}
\centering
    \includegraphics[width=1.0\textwidth]{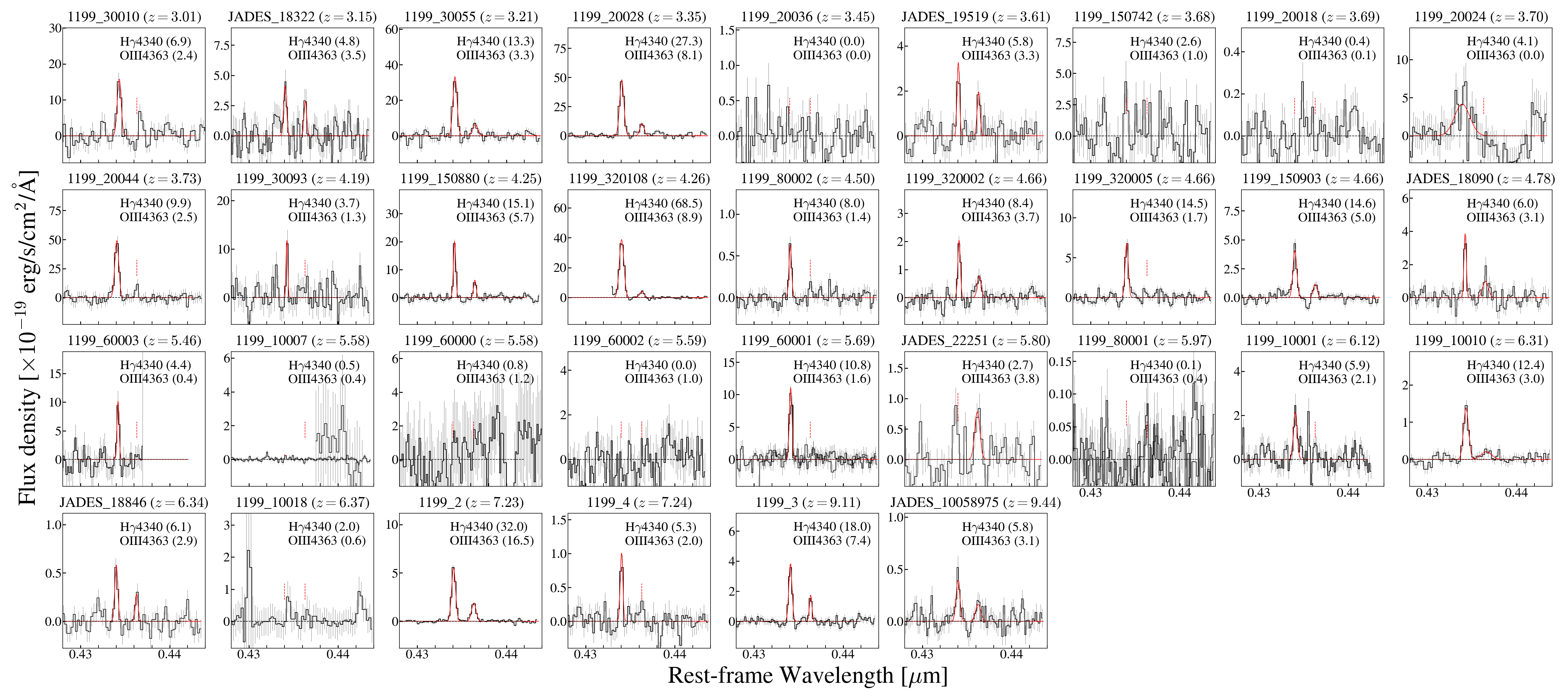}
	\caption{
    NIRSpec MSA spectra 
    of our primary sample galaxies, zoomed in the wavelength region of the \hg\ and \oiiif\ lines. The modelled gaussian is shown when the line is detected at $>3$-$\sigma$ (red solid lines). The detection significance for each line is shown in the label. Note that sources from \citet{nakajima23} are not shown, as we adopt their published flux measurements.
 }
\label{fig:linefit}
\end{figure*}

\subsection{Emission Line Fitting}\label{linefit}
As shown in Fig.~\ref{fig:mosaic}, the slit positions for some sources do not necessarily match in the two grating observations. To accommodate potential systematic errors due to this, we correct for aperture losses. For each grating spectrum of each source, we apply a global scale factor that is derived by comparing the continuum spectrum (with the wavelengths of strong emission lines being masked) to the best-fit spectral template from our spectral energy distribution (SED) analysis (Sec.~\ref{sec:sed}).

We then model the line profile of each emission line of interest in the extracted 1d spectrum with a Gaussian function. Our basic strategy is to define a wavelength window for each line, model the underlying continuum spectrum by a polynomial fit, and subtract it from the 1-d spectrum before the Gaussian fit is carried out. The model includes amplitude, line width, and redshift parameters. However, when multiple lines are located close to each other in wavelength, we fit those simultaneously using a common redshift parameter, so blending can effectively be resolved. In addition, doublet lines (i.e. \oiii$_{\lambda\lambda4959+5007}$ and \nii$_{\lambda\lambda6548+6583}$) are modelled with a fixed line ratio (1:3 and 1:1.7, respectively), with single line width and redshift parameters. The total flux of the doublets estimated in this way remains robust, while we note that the ratio could have a systematic uncertainty $\lesssim10\,\%$ \citep[e.g.,][]{storey00}.

The integrated flux of each line is estimated by summing the corresponding Gaussian component, and the flux error is estimated by summing the error weighted by the amplitude of the Gaussian model in quadrature. 
{The uncertainty associated with the modeled continuum is also added.}
In the following analysis, we adopt flux measurements when the line is detected (signal-to-noise ratio $S/N\geq3$); for those not detected, we adopt the 3-$\sigma$ upper limit.

\subsection{Direct Metallicity Measurements} 
\label{sec:metal}
In general, the line intensity we measure needs to be corrected for dust absorption. We determine the correction on the basis of the ratio of the strongest Balmer lines we measure, thus, \ha/\hb\ is considered before \hb/\hg. We require the Balmer lines to be detected at 5-$\sigma$ in order to apply such correction. In the absence of a high S/N detection of Balmer lines, we apply no correction. Generally, we find the $A_V$ values derived from the Balmer line ratios to be in broad agreement with the SED-derived $A_V$ values, after using the \citet{Calzetti} rescaling factor that accounts for continuum and emission lines dust effects.

The auroral line \oiiif\ provides a reliable method to derive the electron temperature and, from that, the oxygen metallicity by using the \citet{Izotov06} iterative method. The method requires knowledge of the electron density, $n_e$, which we cannot derive directly for (most of) the objects in our sample. In our derivation we assume $n_e = 300\,{\rm cm^{-3}}$. We have estimated the effect of the density uncertainty by assuming also $n_e=100, 1000$ to assess the range of variability. Usually this contribution to the overall uncertainty is small (but see below). The calculations require a well measured {Hydrogen recombination line, ideally \hb, which is close in wavelength to} \oiii$_{\lambda 5007}$. For some objects the line is not measured at S/N$\geq3$; in those cases we estimate \hb\ from another Balmer line using Case B \citep{osterbrock06} and 10,000\,K, assuming no attenuation. For 20018, 60001, 80001, and m39 we estimate \hb\ from \hg, while for 80002 we estimate it from \hd. 

{For objects with \oii\ detection we add the contribution of singly ionized Oxygen to the total Oxygen abundance by estimating the $O^+$ temperature following \citet{Campbell86}, and then follow \citet{Izotov06} to get the $O^+$ abundance for the estimated temperature. 
} These correction are always less than 0.17 and usually less than 0.1 in \logoh\,$+12$. 
{We found similar $O^+$ temperatures  in most cases when we followed \citet{Izotov06} instead of \citet{Campbell86}. However, the polynomial fit that \citet{Izotov06} provide was optimized for the $O^{++}$ temperature of $<2\times10^4$\,K and starts returning unrealistic values at higher temperatures. We therefore decided to adopt \citet{Campbell86} for the $O^+$ temperature estimate.}

We derive Oxygen metallicity for objects with a 3-$\sigma$ or higher S/N for \oiiif. For the remaining objects we adopt a value of \oiiif\ identical to three times the error as an upper limit, which gives us a lower limit to the metallicity. 

{The derived temperatures are reported in Table~\ref{tab:param}. We find that some sources have considerably high temperature, $\simgt2\times10^4$\,K, including 320002 ($T_e=2.6\times10^4$\,K) from GTO1199. Interestingly, all JADES sources are at $\simgt2\times10^4$\,K. While the reason for this is not clear, \citet{laseter23} also reported similarly high temperatures for overlapping sources.}

{We note that the line ratio $I_{4363}/I_{5007}$ become much less sensitive to temperature in high electron density ISM, $n_e\simgt10^4\,{\rm cm^{-3}}$, making our direct method less validated. Indeed, recent observations reported high ($\sim10^5\,{\rm cm^{-3}}$) values in two high-$z$ galaxies \citep{senchyna23,topping24}. However, we consider that such high $n_e$ is unlikely in most of our galaxies. \citet{reddy23} reported an increasing trend of $n_e$ with star formation rate surface density, $\Sigma_{\rm SFR}$. Following their derived slope ($n_e \propto \Sigma_{\rm SFR}^{0.51}$), we find electron density should remain to $<10^4\,{\rm cm^{-3}}$, even for an extreme case with $\Sigma_{\rm SFR}$ of $\sim10^2 M_\odot {\rm kpc^{-3}}$ (cf. \citealt{morishita24}). \citet{isobe23} also found that $n_e$ of their galaxies at $z=6$-10 is $<10^4\,{\rm cm^{-3}}$.}

\begin{figure}
\centering
    \includegraphics[width=0.45\textwidth]{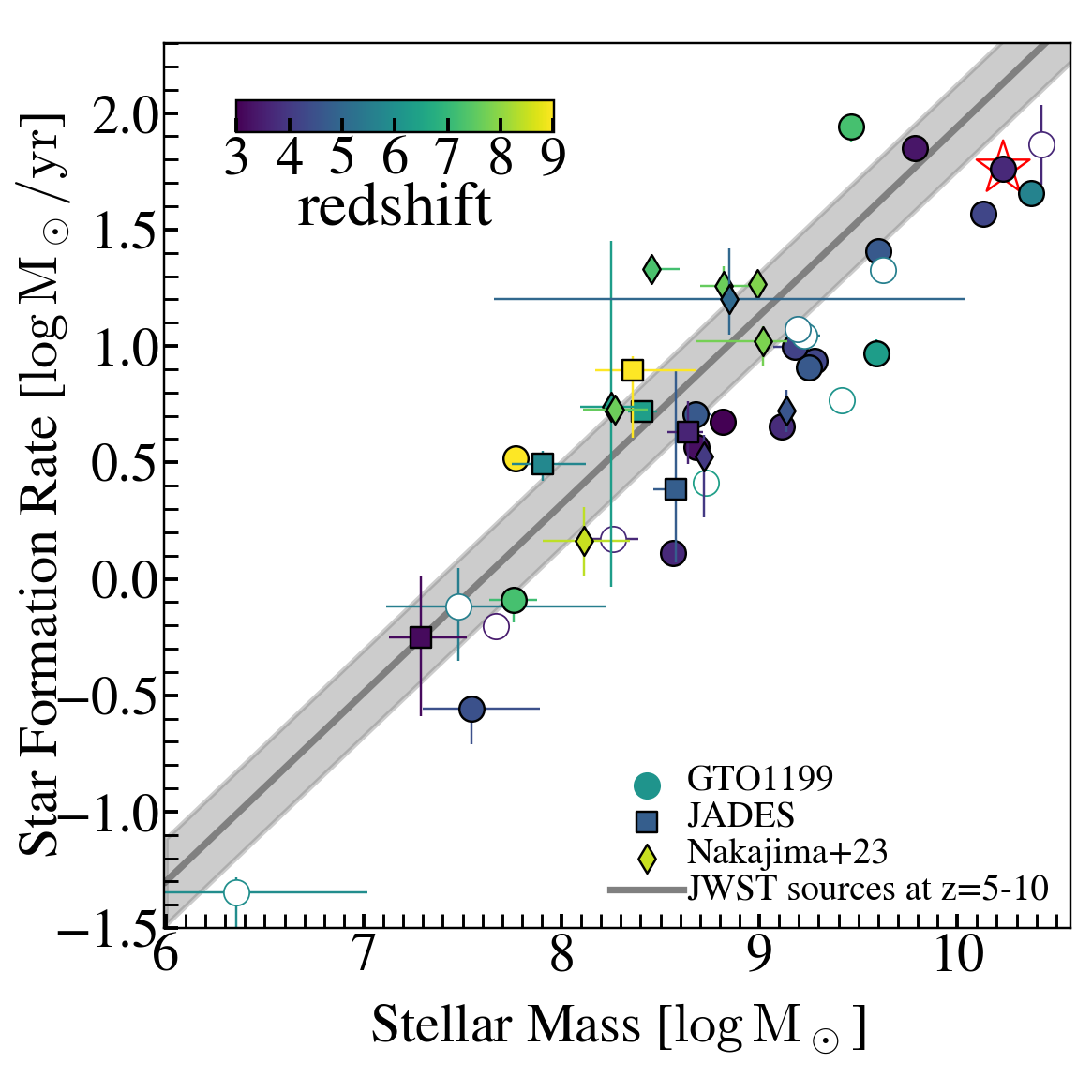}
	\caption{
    Our samples in the stellar mass--star formation rate plane. Symbols are color-coded by redshift. Different symbols represent references (circles for GTO1199, squares JADES, diamonds \citealt{nakajima23}). The linear relation derived for UV-selected galaxies at $5<z<10$ \citep[][solid line]{morishita23b} is shown for comparison. The broad line object, 20024, is marked separately (red star). Those excluded from the mass-metallicity relation analysis (Sec.~\ref{sec:mass-metal}) are shown as open circles.
    }
\label{fig:SFMS}
\end{figure}

\begin{figure*}
\centering
	\includegraphics[width=0.48\textwidth]{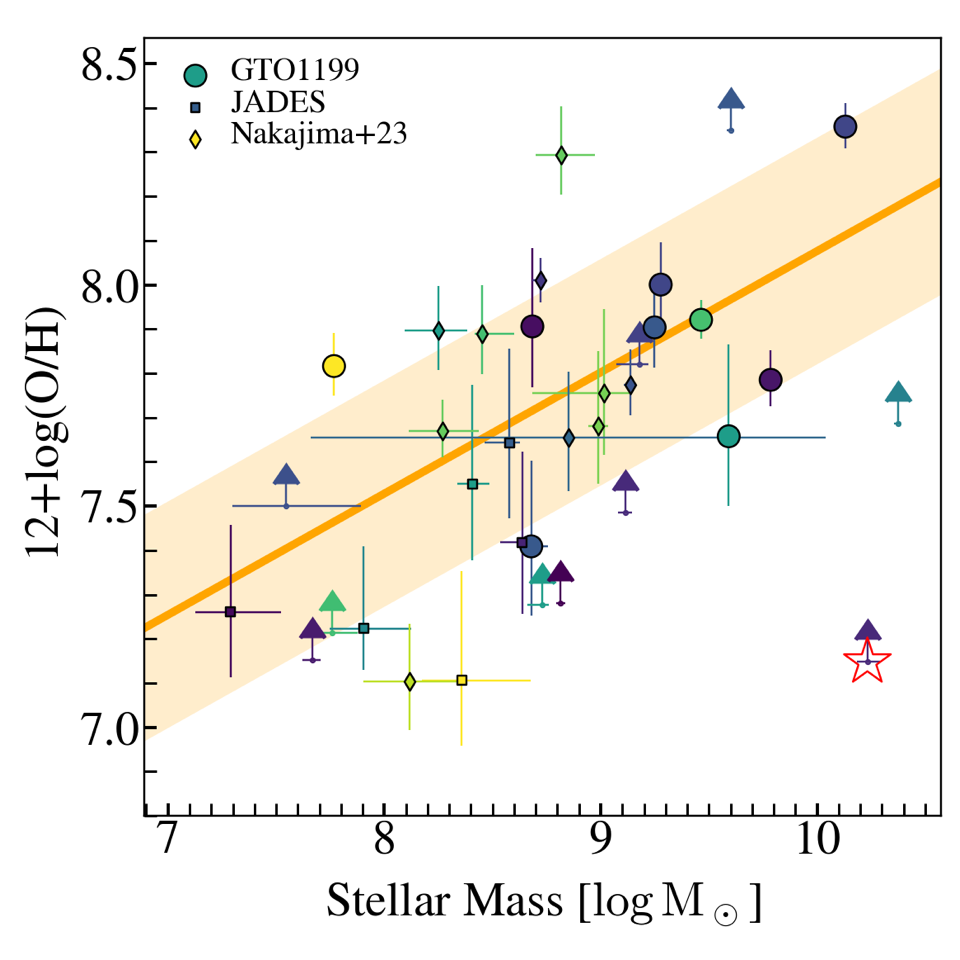}
	\includegraphics[width=0.48\textwidth]{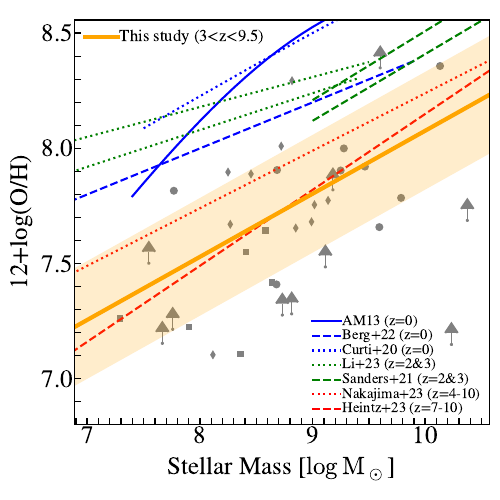}
	\caption{
    (Left): Our sample galaxies in in the 
    stellar mass--metallicity plane. Symbols and colors are the same as in Fig.~\ref{fig:SFMS}. Those with an upper limit in \oiiif\ are shown as a lower limit in \logoh. The best-fit slope to our sample is shown (orange solid line, with hatched regions for the $3$-$\sigma$ range).
    (Right): Same as in the left panel, but with the mass-metallicity relation slopes determined at $z\sim0$ \citep[][]{andrews13,curti20,berg22}, $z\sim2$--3 \citep{sanders21,li23}, and high-$z$ \citep[][]{Heintz2023,nakajima23} in the literature.
    }
\label{fig:MZ}
\end{figure*}

\subsection{Spectral Energy Distribution Analysis}\label{sec:sed}
We infer the spectral energy distribution (SED) of our sample galaxies through SED fitting using the JWST+HST photometric data. We use the SED fitting code {\tt gsf} \citep[ver1.85;][]{morishita19}, which allows flexible determinations of the SED by adopting binned star formation histories (often referred to as {\it non-parametric}). We follow the same procedure as described in \citet{stiavelli23}, but this time without fitting spectroscopic data simultaneously as continuum fluxes of a large fraction of the sources are too faint to be useful.

The star-formation rate of individual galaxies is calculated with the rest-frame UV luminosity ($\sim1600\,\rm{\AA}$) using the determined SED model. The UV luminosity is corrected for dust attenuation using the $\beta_{\rm UV}$ slope, which is measured by using the posterior SED, as in \citet{smit16}:
\begin{equation}
    A_{1600} = 4.43 + 1.99\,\beta_{\rm UV}
\end{equation}
The attenuation corrected UV luminosity is then converted to SFR via the relation in  \citet{kennicutt98}:
\begin{equation}
    {\rm SFR\,[M_\odot\,yr^{-1}]} = 1.4 \times 10^{-28} L_{\rm UV}\,[{\rm erg\,s^{-1}\,Hz^{-1}}].
\end{equation}

\section{Results}\label{sec:res}

\subsection{Spectroscopic Sample from GTO~1199} 
\label{sec:metal}
In Fig.~\ref{fig:mosaic}, we present the cutout images of \nsrc\ sources confirmed to $z>3$ in the GTO1199 observations. The fluxes are reported in Table~\ref{tab:linefluxes}. We note that redshift of all sources is measured by multiple lines, in most cases including the \oiii-doublet.

Among the confirmed sources, \nsrcz\ have $\simgt3\,\sigma$ detection in \oiiif, with ID~3 (JD1) holding the highest redshift, $z=9.11$ \citep[][]{stiavelli23}. The \oiiif\ line of each source is shown in Fig.~\ref{fig:linefit}. Two sources do not have wavelength coverage for \oiiif, due to the wavelength range falling in a detector gap or not being covered by the corresponding grating. 

We note the observed high  \oiiif/\oiii$_{\rm \lambda\lambda 4959+5007}$ ratio for our samples, with the median value of $\sim0.027$ (also Sec.~\ref{sec:pop3}). This is considerably higher compared to those reported in lower-$z$ studies --- for comparison, we find $\sim0.01$ in the Sloan sample for the mass range of $10^8<M_*/M_\odot<10^9$ in the MPA-JHU catalog \footnote{\url{https://www.sdss4.org/dr17/spectro/galaxy_mpajhu/}.}; \citet{sanders16} reported $0.013\pm0.003$ for a galaxy at $z=3.08$. 

Overall, our sample galaxies are characterized by narrow lines. Their location on the stellar mass-SFR plane is shown in Fig.~\ref{fig:SFMS}. The exception is ID~20024, which has broad features in Hydrogen recombination lines. This source, however, shows no significant detection of \oiiif, and thus it is not included in our primary metallicity analysis below. The derived physical properties of our samples are reported in Table~\ref{tab:phys}.

\subsection{Additional \oiiif\ Samples}
In addition, we explore public datasets to expand the sample size of \oiiif sources. We include \nsrcj\ galaxies at $3<z<9.5$ from the JADES program \citep{hainline23,bunker23,eisenstein23} observed in their medium resolution spectroscopy and \nsrcn\ at $4<z<8.6$ from \citet{nakajima23}.

For the JADES sources, we start with the public 1-dimensional spectra and repeat the same line fitting analysis as for the GTO1199 sources. 
The measured fluxes are reported in Table~\ref{tab:linefluxes}. Four of the JADES sources were previously reported in \citet{laseter23}. One of the \citet{laseter23} sources has a solid detection of \oiiif, but has no spectroscopic coverage for the \oiii$_{\lambda\lambda4959+5007}$-doublet (NIRSpec ID 00008083); as such, this source was excluded from our metallicity analysis.

For the sources reported in \citet[][from public JWST fields, ERO \citep{pontoppidan22}, GLASS \citep{treu23}, and CEERS \citep{finkelstein23}]{nakajima23}, we adopt their published measurements for line fluxes and magnification. 

To guarantee a uniform analysis, we derive oxygen abundances using the same method presented in Sec.~\ref{sec:metal}, and stellar mass and star formation rate as in Sec.~\ref{sec:sed}. The derived physical properties are summarized in Table~\ref{tab:phys}. We find overall good agreement of the derived abundances with the previous studies. This provides an independent verification of our method. It also enables a fair internally-consistent comparison with the objects from those studies.

\begin{figure*}
\centering
	\includegraphics[width=0.48\textwidth]{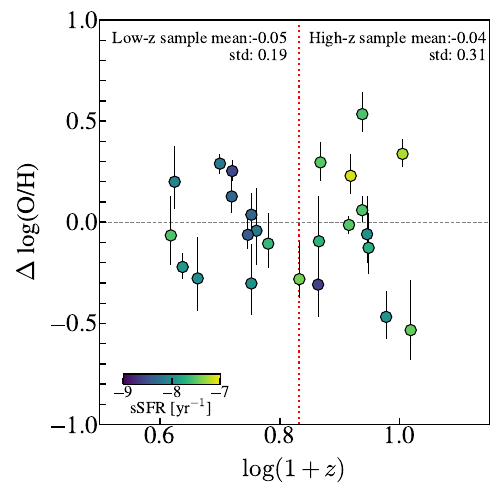}
	\includegraphics[width=0.48\textwidth]{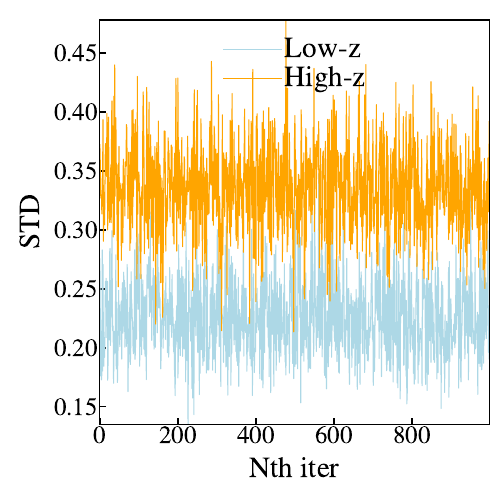}
	\caption{
    (Left):
    Redshift distribution of $\Delta \log\mathrm{(O/H)}$, metallicity offset from the determined mass-metallicity relation in Fig.~\ref{fig:MZ}. The mean and standard deviation are shown for each of low- and high-redshift sub-samples, split at the median redshift (vertical dotted line). Symbols are color-coded by specific star formation rate. Note that the lower limit samples are not shown.  
    (Right): Distributions of standard deviation for low-$z$ (cyan) and high-$z$ samples (orange). Standard deviation was calculated at each iteration with random Gaussian noise for the observed uncertainties being added.
    }
\label{fig:delMZ}
\end{figure*}

\subsection{Stellar Mass -- Metallicity Distribution of Galaxies at $3<z<9.5$}\label{sec:mass-metal}

In the left panel of Fig.~\ref{fig:MZ}, we show the distribution of our sample galaxies in the stellar mass-metallicity plane. Despite a wide redshift range, our samples overall distribute around the linear relation reported in the literature, derived for $z>3$ galaxies \citep{heintz23,nakajima23}. 

We investigate the distribution of our samples by using the following formulae:
\begin{equation}\label{eq:massz}
\begin{split}
12 + \log {\rm (O/H)} &  = \alpha \log (M_*/M_{\rm 0})\\ & + B(z) + N(0, \sigma_{\rm log MZR}),
\end{split}
\end{equation} 
\begin{equation}\label{eq:B}
B(z) = \beta_z + \alpha_z \log (1 + z),
\end{equation}
where $N(0, \sigma_{\rm log MZR})$ represents a normal distribution with $\sigma_{\rm log MZR}$ being the intrinsic scatter around the relationship. For the calculation of posterior distributions, we employ the Markov chain Monte Carlo method by using {\tt emcee} \citep{foreman14}. The likelihood function is defined as:
\begin{equation}\label{eq:emcee}
ln \mathcal{L} = -0.5 \left(\sum_{i} (\Delta_i^2 + ln \left(2 \pi \sigma_i^2)\right) - 2 \chi_{lim} \right),
\end{equation}
where $i$ runs through detected sources (i.e. S/N \oiiif\,$>3$). $\Delta_{i}$ is the normalized residual: 
\begin{equation}\label{eq:emcee2}
\Delta_{i} = (\log {\rm (O/H)}_{\rm model} - \log {\rm (O/H)}_i) / \sigma_i.
\end{equation}
The term of $\chi_{lim}$ accounts for the likelihood of {\it lower}-limit sources \citep[also,][]{sawicki12}:
\begin{equation}\label{eq:emcee3}
\chi_{lim} = \sum_{j} ln \left(\sqrt{\frac{\pi}{2}}\sigma_j\bigl(1+{\rm erf}(\frac{\log {\rm (O/H)}_{\rm model} - \sigma_j}{\sqrt{2}\sigma_j})\bigr)\right)
\end{equation}
where $j$ runs through lower-limit sources, $\sigma_j$ lower limit on \logoh, and erf the error function.

We adopt the median mass of the sample as the pivot mass, $M_{\rm 0}$. From the regression analysis, we find the slope $\alpha = 0.27\pm0.10$, similar to \citet{nakajima23}. We find a slightly smaller offset ($\sim-0.15$\,dex) from the \citet{nakajima23}'s relation, but the scatter around our determined relationship is considerably large ($\sigma_{\rm \log MZR}=0.26\pm0.05$). 

The derived intercept is $\sim0.4$--0.7\,dex lower than that in the local relation at the pivot mass, $\log M_0 = 8.8$ \citep[e.g.,][among those adopt the direct \oiiif\ method]{andrews13,curti20,berg22}. This indicates evolution of the relation, from $z\sim3$--9 to $z\sim0$, supporting the previous conclusion reported in various studies at similar redshifts \citep[e.g.,][]{nakajima23,heintz23}.

On the other hand, redshift evolution of metallicity {among our sample galaxies} is not observed. The relevant parameter, $\alpha_z$, is consistent with no evolution (i.e. $\alpha_z = 0.0\pm0.5$) in our regression analysis. 
Indeed, we find that our samples are distributed in a fairly wide range of metallicity regardless of redshift. For example, the two highest redshift sources (JD1 and 10058975) have distinct values (\logoh\,$+12=7.81$ and 7.08). ERO~6355 at $z=7.655$ is one of the most enriched galaxies in the combined sample, with \logoh\,$+12=8.29$. We investigate the observed scatter in Sec.~\ref{sec:sct}.

It has been known that at lower redshifts mass-metallicity relations have dependency on a third parameter, such as SFR \citep[e.g.,][]{andrews13} and gravitational potential \citep[e.g.,][]{sanchez-menguiano23}. However, none of those quantities offers significant reduction in the observed scatter. For example, with $M_* / {\rm SFR}^{0.66}$ \citep[i.e. the fundamental-metallicity;][]{andrews13,curti20,li23}, we only find a comparable intrinsic scatter, $\sigma_{\rm \log MZR}=0.28$.

{
To understand the observed scatter, we investigate $\Delta$\,\logoh, the deviation of metallicity from the determined relation, by correlating it with a third parameter. We find that when correlated with sSFR, the Pearson's $r$ value is $0.10$ with a corresponding $p$-value of $0.64$, indicating a lack of correlation. Similarly, we do not observe significant trends with other parameters, including SFR ($r=0.07$ and $p=0.74$), $\Sigma_{\rm SFR}$ ($-0.26, 0.21$), gravitational potential ($-0.26, 0.21$), stellar mass surface density ($-0.26, 0.21$), light-weighted age ($-0.26, 0.21$), and UV slope $\beta$ ($0.07,0.74$). 
}

\section{Discussion}\label{sec:disc}

\subsection{On the presence of the Mass-Metallicity relation}\label{sec:sct}
We have observed correlation between stellar mass and metallicity for our sample galaxies. Despite the large scatter, the presence of a positive slope indicates that stellar mass formation plays a critical role in enrichment process already at these redshifts.

The observed slope of our galaxies is offset from the local relation for $\sim0.4$--$0.7$\,dex, depending on the local reference. A similar offset is reported in other studies at similar redshifts \citep[e.g.,][]{heintz23}, and at $z\sim3$ \citep[e.g.,][]{sanders20,curti23}, the lower bound of our redshift range. In contrast, we find no clear evidence of redshift evolution among our sample. We note that the time difference is much smaller at $3<z<9$ ($\sim1.5$\,Gyr; whereas $\sim10$\,Gyr from $z\sim3$ to $0$), while some studies reported the presence of weak evolution using a sample of calibrated metallicity \citep[e.g.,][]{curti23}. 

One possible interpretation is that what regulates metallicity changes around $z\sim2$--3, from bursty to more continuous star formation \citep[e.g.,][]{Asada24}. At higher $z$, the relevant ``clock'' is driven by mass, rather than the specific redshift, and galaxies will enrich until perhaps an outflow quenches star formation or ejects metals. At decreasing redshift, the star formation history becomes more important, as galaxies begin retaining metals or perhaps accreting metal rich gas so their metallicity can be higher for a given mass. It remains surprising that slopes remain similar across the cosmic time.

The absence of redshift evolution at $z\gtrsim5$ is reported in numerical simulations. \citet{ucci23} reported a decrease of metallicity for $\logm\sim7$ galaxy systems at decreasing redshift. They attribute this to various factors, including the differential impact of feedback in different halo mass systems at low and high redshifts. The impact by feedback becomes smaller at a larger mass system, making the ``observed" evolution insignificant. Similarly weak/no redshift evolution of the relation is also reported in other numerical work \citep[e.g.,][]{ma16,langan20}.

\begin{figure*}
\centering
	\includegraphics[width=0.48\textwidth]{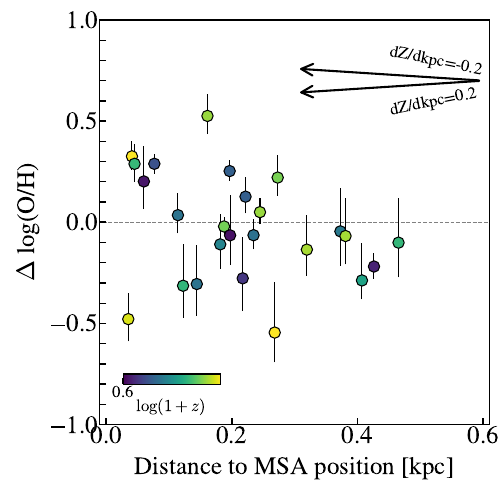}
	\includegraphics[width=0.48\textwidth]{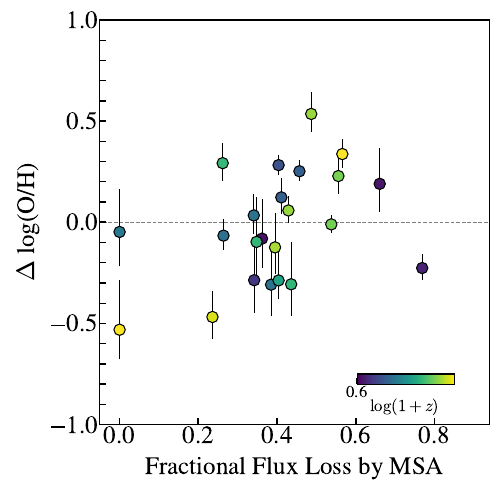}
	\caption{
Same as Fig.~\ref{fig:delMZ}, but with different quantities for the $x$-axis. (Left): Distance from the MSA shutter position to the light-weighted center of the target galaxy. The distribution of our samples does not reveal a clear trend with the offset amplitude. The deviation from the mean relation cannot be explained by the offset, even under the presence of steep metallicity gradient ($0.2$\,dex/kpc and $-0.2$\,dex/kpc; arrows). (Right): Fractional flux loss of MSA (i.e. flux that falls out of the MSA slit over the total flux). The fraction flux loss estimated in F444W is shown here.
    }
\label{fig:msaoff}
\end{figure*}

\subsection{On the scatter around the Mass-Metallicity relation}\label{sec:sct}
We have found a considerably large scatter around the determined relation. Our regression analysis above concluded that the scatter cannot be attributed to a third quantity, including the redshift dependence of the relation. 
For comparison, at $z\sim0$ the scatter is found to be $\sim0.1$ around the fundamental mass-metallicity relation at the same mass range \citep{curti20}.

{While we discuss observational systematics that may attribute to the scatter in Section~\ref{sec:sys}, here we focus on potential attributes arising from the nature of our sample galaxies.
The large scatter could be interpreted} that metal enrichment process occurs rapidly, but also susceptible to mechanisms that dilute the enrichment over a similar time scale. Such inverse processes are expected to be less significant at lower redshift, where a stronger gravitational potential aids retaining the enriched gas within the galaxy system against feedback processes. We note that in the stellar mass range of interest here, the halo mass of the system increases as redshift decreases until $z\sim1$ (and then it decreases toward $z\sim0$; e.g., \citealt[][]{behroozi19}). 

Our data supports the interpretation. In Fig.~\ref{fig:delMZ}, we show $\Delta$\logoh\ as a function of redshift. Clearly, the amplitude of the scatter increases at increasing redshift. The standard deviation for the lower-half and higher-half redshift samples are $0.17\pm0.04$ and $0.28\pm0.07$, respectively. {Since the sample size is not large, we adopt a simple Gaussian estimator and bootstrap method to derive the quoted uncertainty \citep[][also \citealt{morishita23b}]{beers90}. We have also confirmed the result through an iterative approach, where the standard deviation with scatter added to the metallicity measurements at each iteration (Fig.~\ref{fig:delMZ}, right panel).}

Notably, high-$z$ galaxies are intense in star formation. In \citet{morishita23b}, we reported that star formation rate surface density, $\Sigma_{\rm SFR}$, of $z>5$ galaxies is $\sim10$--$100$ times higher than $z\sim2$ samples, driven both by compact morphology and high star formation rate. The decrease of scatter amplitude can be explained such that interplay between the potential well and feedback process becomes more constrained at decreasing redshift. 

We remind the absence of clear correlation between $\Delta$\,\logoh\ and sSFR in our sample. The lack of correlation with sSFR could simply be due to the fact that chemical enrichment may not necessarily proceed in the same time scale for star formation at these redshifts ($\simlt200$\,Myr; \citealt{faisst24}). 

Alternatively, an increased fraction of merger-driven evolution would also explain the observed trend of the scatter with redshift. As redshift increases, so does the merger rate \citep[][also \citealt{hsiao23,boyett24,dalmasso24}]{fakhouri2008}, and thus galaxy assembly is progressively more driven by mergers than smooth accretion, {though the actual merger rate for low-mass galaxies at these redshifts is yet to be observationally constrained.} As a result, the system's chemical enrichment is more affected by independent building blocks rather than by smooth accretion of material and self-recycling.

Simulations find a different conclusion. \citet{ucci23} reported a dependence of the relationship on specific star formation rate at $z>5$, in a way that a lower metallicity galaxy has a higher sSFR. Their observed trend is interpreted as the strong star formation could blow the enriched gas away from the system, while pristine gas may still feed the galaxy system. We do not observe this trend in our sample. {A larger sample may be needed to confirm the predictions.}

\subsection{Potential Systematic Uncertainties}\label{sec:sys}
Slit-spectroscopy observations have two major limitations, namely the position of slit regarding the galaxy center and aperture loss due to the limited slit size. Since the mass-metallicity relation concerns the global metallicity of the system, those systematic uncertainties need to be addressed. Those effects are particularly the case for JWST for its design of MSA masks, its resolving power, and its relatively small size of the slit ($0.\!''2\times0.\!''46$). Despite its critical importance, the literature has not yet engaged in a comprehensive discussion on these systematic issues.

In our observations, we find that several targets show a small offset in the MSA slit position from its light center (Fig.~\ref{fig:mosaic}). This is partially due to our selection of ``Entire Open Shutter Area" configuration in APT, chosen to maximize multiplexity, leading to an offset of $\sim 0.25$\,arcsec at maximum. Such an offset would have no impact to mass-metallicity relation if the metallicity measured at the position represents the entire system. However, this is not necessarily the case. Studies have found a pronounced gradient in gas-phase metallicity in active galaxies, up to $z\sim3$ \citep[e.g.,][]{belfiore17,bresolin19,wang20}. 

We investigate this by examining the deviation in metallicity from the mean relationship in relation to the offset distance (Fig.~\ref{fig:msaoff}, left). Our analysis reveals no clear indication that the offset amplitude is the primary cause for the observed scatter. In fact, the observed scatter amplitude surpasses what can be accounted for by an extreme gradient reported in the literature ($\pm0.2$\,dex/kpc). {Recent observations revealed negative metallicity gradient in galaxies of similar redshifts \citep{arribas23,vallini24}, while the scatter is likely more significant due to shorter mixing timescales of gas-phase metallicity \citep[e.g.,][]{venturi24}.}

We then investigate the impact by aperture loss. We here measure fractional aperture loss ($\Delta f_{\rm MSA}$), i.e. flux that falls out of the slit over the total flux of the target, in the F444W image. The right panel of Figure~\ref{fig:msaoff} shows the distribution. This quantity ranges from $\sim0$ (i.e. most flux is within the slit) to $0.8$ ($20\,\%$ is within the slit) for our sample, with the median value $0.4$. Interestingly, three sources that are least affected by the loss ($\Delta f_{\rm MSA}<0.25$) have metallicity below the mean relation. On the other hand, at the high-loss end ($\Delta f_{\rm MSA}\simgt0.5$), four (3, 30055, ERO\_6355, GLASS\_10021) out of six are located above the mean metallicity relation. 

Despite the observed signal being statistically weak, this reveals an important implication. Since the MSA slit position is, in default, configured to be near the light center of the target, the inferred metallicity would be biased to the central part, where metallicity is likely most enriched. The effect would become significant when there exists a negative gradient {\it and} the aperture loss is large (i.e. large-sized target), leading to an overestimated metallicity. In other words, the four sources of high $\Delta f_{\rm MSA}$ may be observed to be metal rich {\it because} their low-metallicity part might have fallen out of the slit. Qualitatively, the impact should remain $\simlt0.2$\,dex reduction in metallicity at most for the typical apparent size of our galaxies assuming a gradient of $-0.2$\,dex/kpc. A more realistic estimate could be achieved upon the consideration of, e.g., the actual light profile and metallicity gradient, which is beyond the scope of this study.

We also repeat the regression analysis but this time with 24 galaxies with those fractional flux loss $<0.5$. We do not see decrease in the scatter ($\sigma_{\rm \log MZR}=0.27$; Table~\ref{tab:param}). We find positive redshift evolution of the intercept, but the significance remains too tentative to confirm the trend ($\alpha_{z}=-0.3\pm0.7$).

\begin{figure}
\centering
	\includegraphics[width=0.23\textwidth]{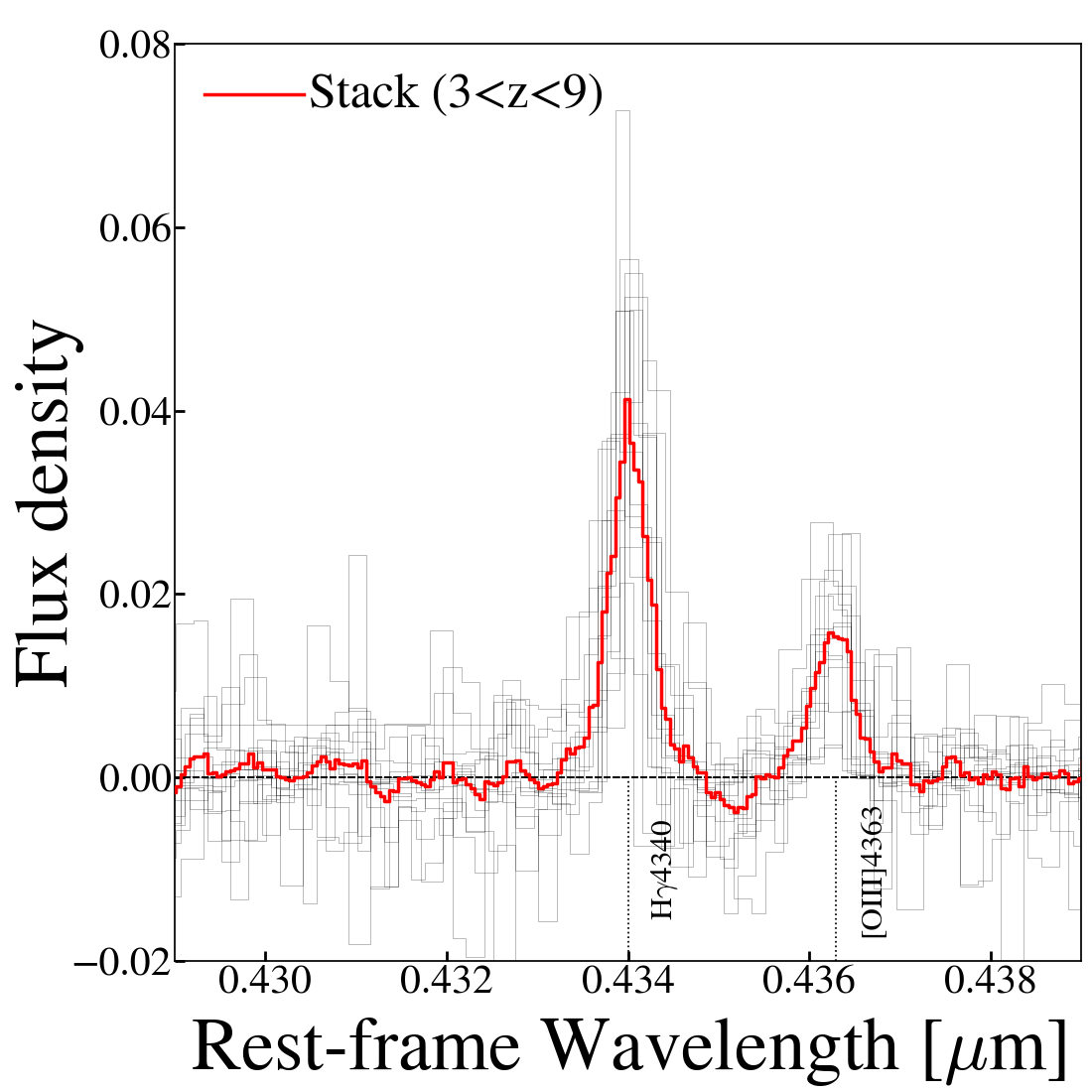}
	\includegraphics[width=0.23\textwidth]{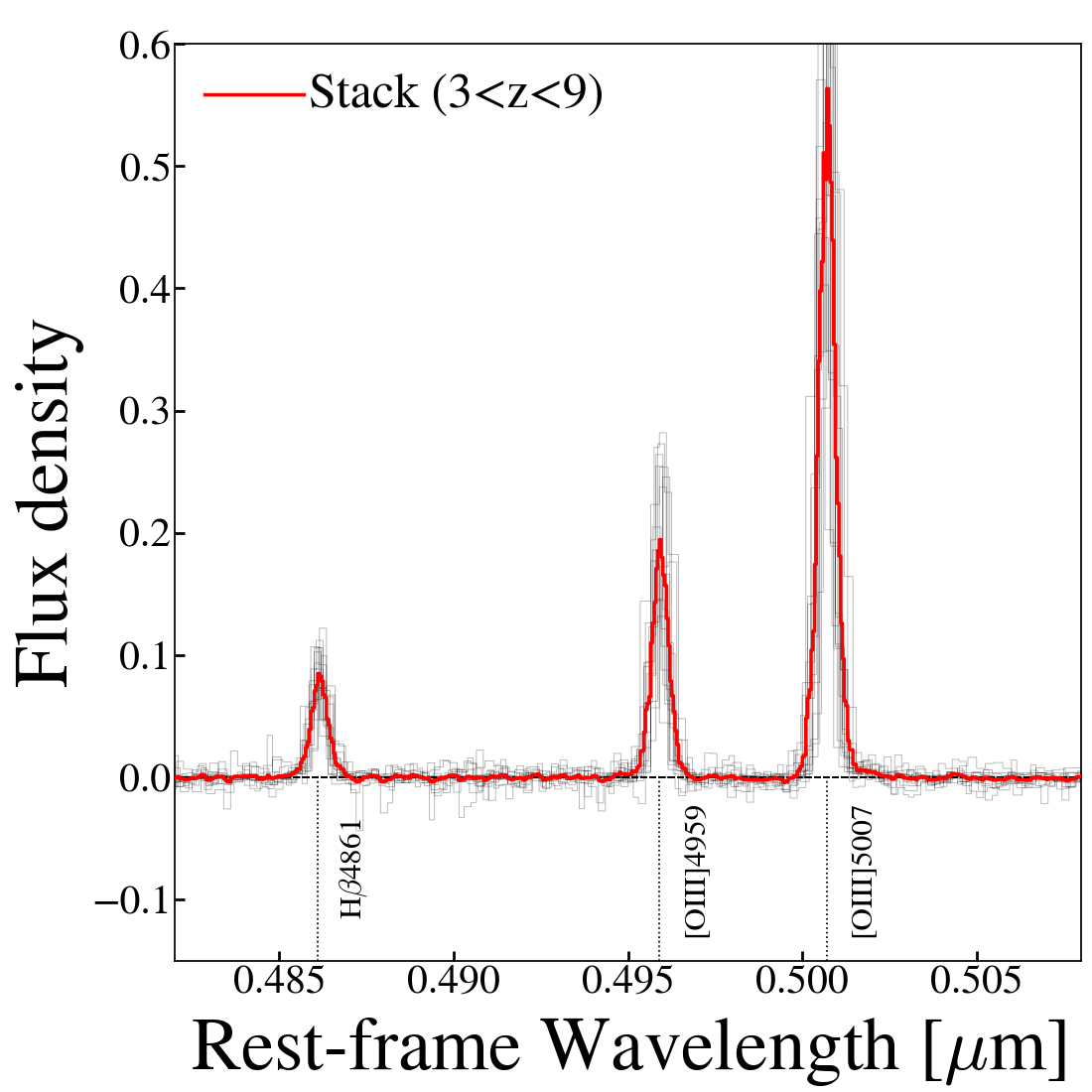}
	\includegraphics[width=0.23\textwidth]{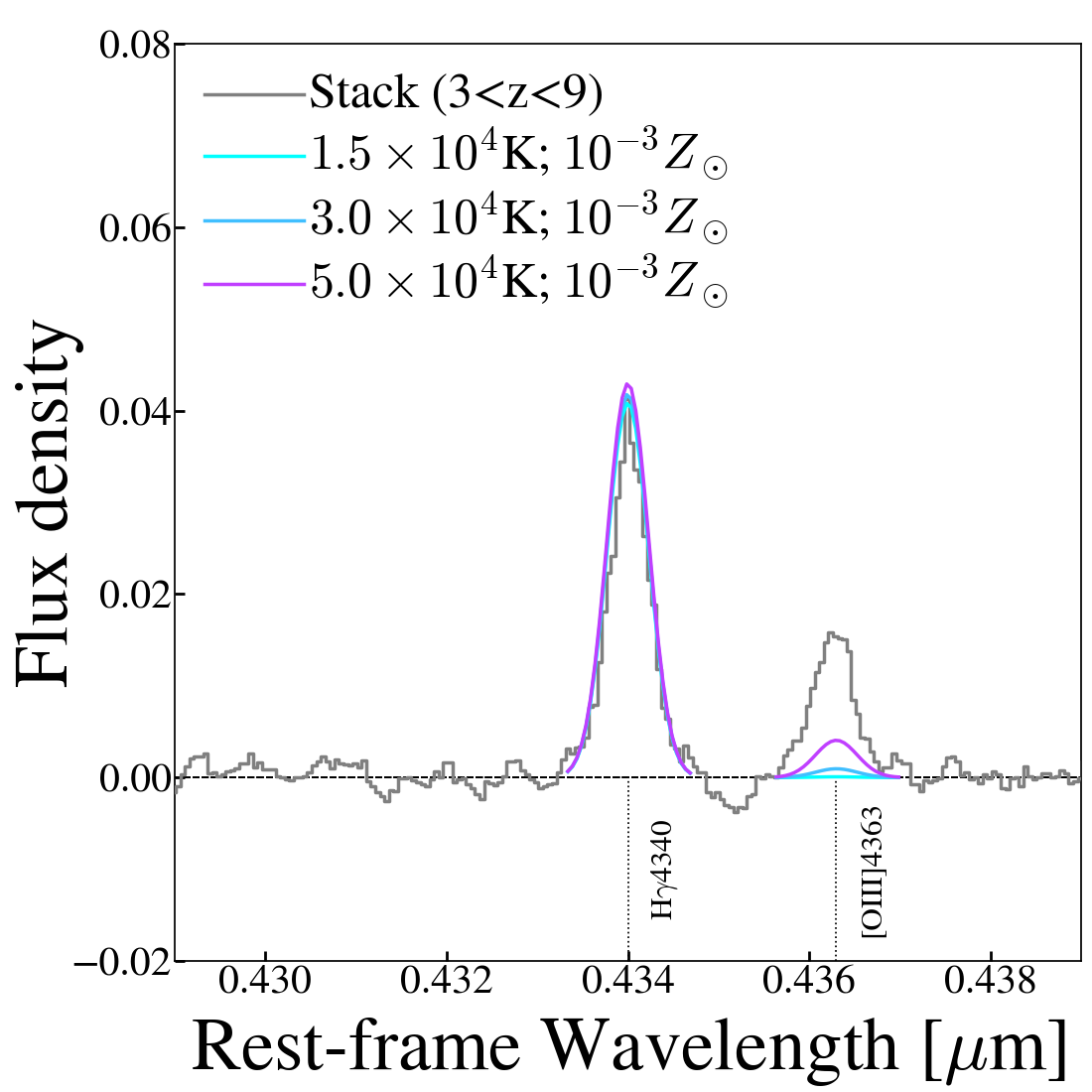}
	\includegraphics[width=0.23\textwidth]{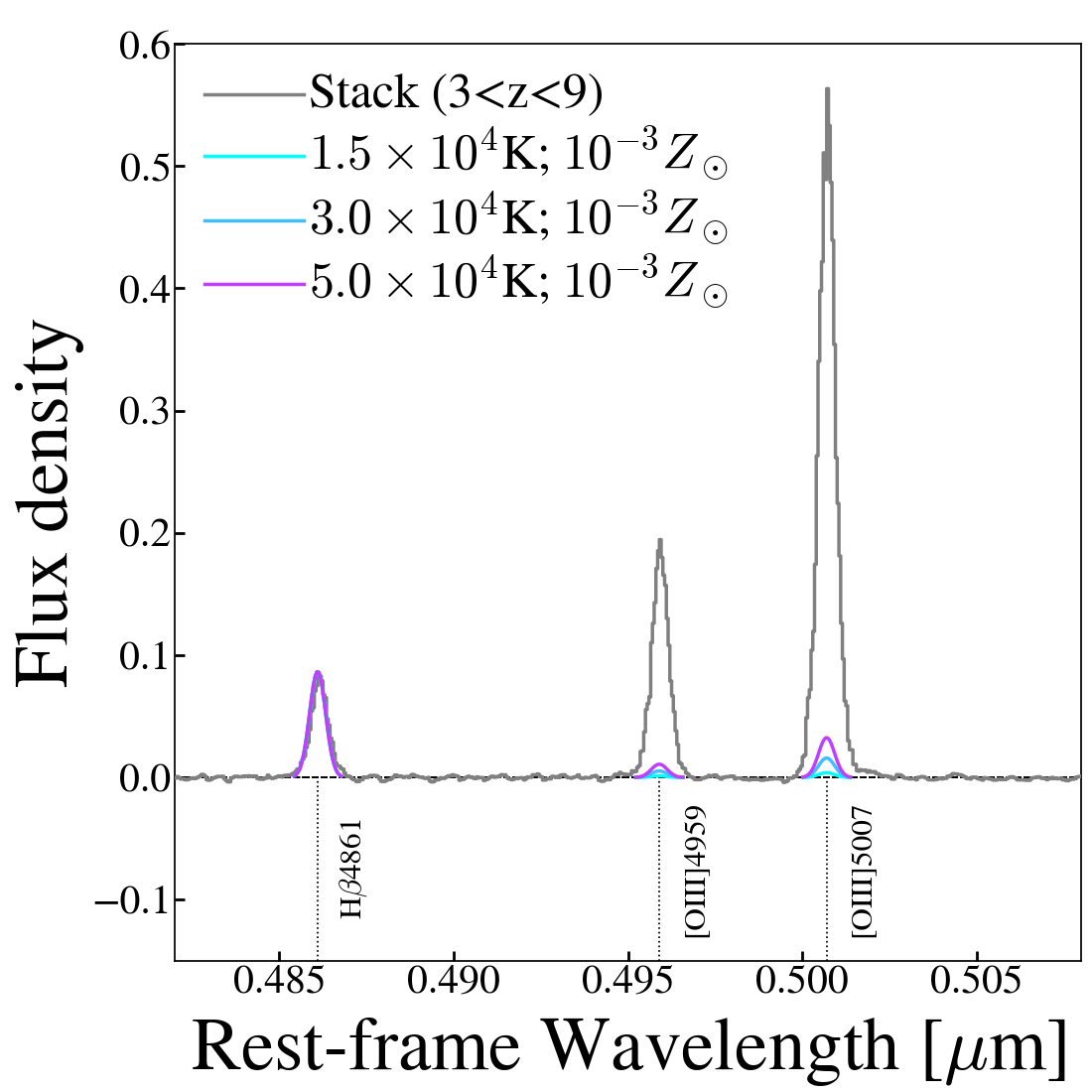}
	\caption{
    (Top panels): Stacked spectrum of the sample galaxies that have \oiiif\ detected (red lines). Individual spectra are shown in the background (gray lines). All spectra were subtracted continuum and normalized by the integrated \hb\ line flux before being stacked.
    (Bottom panels): Same stacked spectra are shown (gray), along with three metal-poor models of for different temperatures ($1.5, 3$, and $5\times10^4$\,K, with $Z=10^{-3}\,Z_\odot$).
    }
\label{fig:stack}
\end{figure}
\subsection{Toward the identification of Pop-III galaxies}\label{sec:pop3}
Through the robust determination of oxygen abundance, our data reveal the distribution of early galaxies in the mass-metallicity plane. Despite probing a wide range of both redshift and stellar mass, our galaxies turn to be moderately enriched in metallicity, \logoh\,$+12\simgt7$. For comparison, the protoype of low-metallicity galaxy in the local Universe, I Zw 18, has \logoh\,$+12\ = 7.17$ \citep{Izotov99}. The lowest mass galaxy among our mass-metallicity relation sample, with $\logm = 7.3$ (ID 18322), has oxygen abundance of \logoh\,$+12\sim7.3$.
This points us to the need for further exploration toward lower mass, $\logm\ll7$, even at such an early time of the universe, to find a less enriched galaxy system.

In Fig.~\ref{fig:stack}, we show the stacked spectrum of our samples. Given that the temperature becomes high in low-metal ISM, the line ratio \oiiif/\oiii$_{\rm \lambda\lambda 4959+5007}$ also becomes larger, leaving us a chance to apply the direct-$T_e$ method down to a considerably low-metallicity regime. Shown along with our stacked spectrum are a few predictions for the lines. Depending on the electron temperature, the \oiiif\ line can be detected down to \logoh$+12\approx 6$ for the sensitivity enabled by JWST. 

Recently, \citet{vanzella23} reported an upper limit of \logoh$+12<6.3$ on a low-mass star complex ($\simlt10^4\,M_\odot$), through calibrated strong line methods. The system is a part of lensed arcs strongly amplified by the foreground galaxy cluster. Sub-scale observations, with an assist of strong gravitational lensing \citep[also,][]{adamo24}, is a path toward the identification of pristine systems \citep[also, e.g.,][]{bovill24}. 

Lastly, we note that the MSA slits of our NIRSpec observations were placed near the galaxy light-center, as the default option. This is presumably the case for many other MSA programs. Given that the galaxy center is likely the place where metal enrichment is more advanced, spectroscopy of off-centered regions may also help such endeavor.

\section{Summary}\label{sec:sum}
In this study, we have reported \nsrc\ galaxies at $3<z<9.1$ in the sightline of the cluster of galaxies MACS~J1149. We detected the auroral \oiiif\ line in \nsrcz\ sources, which allows robust oxygen abundance determination via the direct-$T_e$ method. By combining with \nsrcext\ sources from publicly available data, we found that the mass-metallicity relation exists at the redshift range of interest, with $\sim0.4$--0.7\,dex offset down below the local relation. We found that the scatter of our sources around the relation is not small. Interestingly, the deviation becomes smaller as redshift decreases, which is indirect evidence that the interplay between gravitational potential well and feedback intensity converge. We find no clear evidence of redshift evolution in our sample. While seen in previous lower-$z$ observations and predicted in simulations, we did not find the third parameter that can effectively reduce the observed scatter. Despite reaching to a considerably low-mass with an assist of strong lensing, our samples exhibit a relatively enriched ISM, indicating that the exploration of lower-mass galaxies or sub-galactic scale is needed for the search of less enriched populations, even at the redshift range explored in this study.

We investigated potential systematic impacts arising from the constraints of JWST's MSA observations. It is important to bear in mind that the metallicity measured using MSA slits may not necessarily represent the ``global" metallicity, which is the quantity of primary concern in mass-metallicity relation analyses. This effect is likely to be more pronounced at lower redshifts or in lensing fields, where the anticipated flux loss is greater. 


\begin{deluxetable*}{ccccccccccc}
\tabletypesize{\footnotesize}
\tablecolumns{11}
\tablewidth{0pt} 
\tablecaption{
Rest-frame optical line flux measurements of $3<z<9.5$ galaxies from the GTO 1199 program.
}
\tablehead{
\colhead{ID} & \colhead{[OII]$_{3727}$} & \colhead{[NeIII]$_{3869}$} & \colhead{[NeIII]$_{3968}$} & \colhead{H$\delta_{4102}$} & \colhead{H$\gamma_{4340}$} & \colhead{[OIII]$_{4363}$} & \colhead{H$\beta_{4861}$} & \colhead{[OIII]$_{4959+5007}$} & \colhead{[NII]$_{6548+6583}$} & \colhead{H$\alpha_{6563}$}
}
\startdata
\cutinhead{GTO~1199 Sample}
30010 & -- & -- & -- & -- & {\color{azure}$34.5 \pm 15.0$} & {\color{azure}$<11.7$} & {\color{azure}$63.2 \pm 12.5$} & {\color{azure}$330.3 \pm 54.7$} & {\color{azure}$<71.6$} & {\color{azure}$147.6 \pm 59.1$}\\
30055 & -- & -- & -- & {\color{azure}$47.5 \pm 28.2$} & {\color{azure}$79.3 \pm 17.9$} & {\color{azure}$18.8 \pm 16.9$} & {\color{azure}$118.4 \pm 16.1$} & {\color{azure}$1358.7 \pm 102.1$} & {\color{azure}$<19.3$} & {\color{azure}$438.3 \pm 20.2$}\\
20028 & -- & -- & {\color{azure}$63.0 \pm 19.2$} & {\color{azure}$59.6 \pm 19.0$} & {\color{azure}$123.6 \pm 13.6$} & {\color{azure}$26.4 \pm 9.8$} & {\color{azure}$267.1 \pm 20.5$} & {\color{azure}$2061.2 \pm 90.8$} & {\color{azure}$85.1 \pm 32.9$} & {\color{azure}$954.4 \pm 26.2$}\\
20036 & -- & {\color{azure}$<6.5$} & {\color{azure}$<4.6$} & {\color{azure}$3.7 \pm 3.5$} & {\color{azure}$<2.8$} & {\color{azure}$<2.6$} & {\color{azure}$1.5 \pm 1.1$} & {\color{azure}$12.3 \pm 5.2$} & {\color{azure}$<1.9$} & {\color{azure}$4.7 \pm 1.1$}\\
150742 & {\color{azure}$92.1 \pm 31.2$} & {\color{azure}$<19.7$} & {\color{azure}$<13.2$} & {\color{azure}$<11.4$} & {\color{azure}$<9.3$} & {\color{azure}$<14.4$} & {\color{azure}$31.8 \pm 11.4$} & {\color{azure}$155.5 \pm 49.7$} & {\color{alizarin}$79.8 \pm 26.1$} & {\color{alizarin}$219.4 \pm 12.5$}\\
20018 & {\color{azure}$<1.4$} & {\color{azure}$<0.8$} & {\color{azure}$<0.9$} & {\color{azure}$<0.8$} & {\color{azure}$<0.8$} & {\color{azure}$<0.9$} & -- & {\color{azure}$12.5 \pm 2.6$} & {\color{alizarin}$<3.4$} & {\color{alizarin}$126.2 \pm 7.7$}\\
20014 & -- & -- & -- & -- & -- & -- & -- & -- & {\color{alizarin}$<1.4$} & {\color{alizarin}$78.5 \pm 4.1$}\\
20024 & {\color{azure}$104.3 \pm 19.1$} & {\color{azure}$23.8 \pm 19.4$} & {\color{azure}$<27.2$} & {\color{azure}$47.6 \pm 24.7$} & {\color{azure}$41.3 \pm 30.2$} & {\color{azure}$<20.1$} & {\color{azure}$183.3 \pm 30.1$} & {\color{azure}$345.3 \pm 83.4$} & {\color{azure}$436.8 \pm 75.7$} & {\color{azure}$752.6 \pm 37.3$}\\
20044 & {\color{azure}$<48.3$} & {\color{azure}$67.9 \pm 38.8$} & {\color{azure}$<85.7$} & {\color{azure}$60.7 \pm 24.8$} & {\color{azure}$127.3 \pm 38.7$} & {\color{azure}$<24.2$} & {\color{azure}$230.5 \pm 21.8$} & {\color{azure}$1067.2 \pm 84.2$} & -- & --\\
30093 & {\color{azure}$100.0 \pm 35.1$} & {\color{azure}$17.3 \pm 17.1$} & {\color{azure}$<29.7$} & {\color{azure}$<13.4$} & {\color{azure}$14.5 \pm 11.6$} & {\color{azure}$<12.7$} & {\color{azure}$47.6 \pm 15.5$} & {\color{azure}$332.7 \pm 67.7$} & {\color{alizarin}$<13.5$} & {\color{alizarin}$160.7 \pm 28.1$}\\
150880 & {\color{azure}$88.1 \pm 12.3$} & {\color{azure}$52.0 \pm 8.5$} & {\color{azure}$26.0 \pm 9.4$} & {\color{azure}$24.6 \pm 7.7$} & {\color{azure}$34.4 \pm 6.8$} & {\color{azure}$11.9 \pm 6.2$} & {\color{azure}$87.6 \pm 8.2$} & {\color{azure}$898.7 \pm 47.5$} & {\color{alizarin}$23.3 \pm 11.0$} & {\color{alizarin}$286.9 \pm 9.0$}\\
320108 & {\color{azure}$495.6 \pm 19.6$} & {\color{azure}$137.4 \pm 12.0$} & {\color{azure}$74.4 \pm 10.3$} & -- & {\color{azure}$128.9 \pm 5.6$} & {\color{azure}$13.3 \pm 4.5$} & {\color{azure}$321.8 \pm 11.9$} & {\color{azure}$2555.7 \pm 61.7$} & -- & --\\
80002 & {\color{azure}$<0.9$} & {\color{azure}$1.7 \pm 1.1$} & {\color{azure}$<1.5$} & {\color{azure}$1.9 \pm 1.0$} & {\color{azure}$1.3 \pm 0.5$} & {\color{azure}$<0.5$} & -- & {\color{azure}$12.4 \pm 3.0$} & -- & --\\
320002 & {\color{azure}$8.8 \pm 2.0$} & {\color{azure}$4.1 \pm 1.6$} & {\color{azure}$3.5 \pm 1.9$} & {\color{azure}$2.5 \pm 1.9$} & {\color{azure}$4.0 \pm 1.4$} & {\color{azure}$2.4 \pm 1.9$} & {\color{azure}$9.5 \pm 3.1$} & {\color{azure}$78.0 \pm 15.3$} & {\color{alizarin}$<0.7$} & {\color{alizarin}$37.6 \pm 2.7$}\\
320005 & {\color{azure}$82.6 \pm 8.3$} & {\color{azure}$18.7 \pm 5.2$} & {\color{azure}$8.1 \pm 5.0$} & {\color{azure}$6.0 \pm 3.9$} & {\color{azure}$16.7 \pm 3.5$} & {\color{azure}$<3.7$} & {\color{azure}$42.5 \pm 5.6$} & {\color{azure}$363.0 \pm 28.2$} & -- & --\\
150903 & {\color{azure}$21.2 \pm 4.8$} & {\color{azure}$12.6 \pm 3.9$} & {\color{azure}$5.6 \pm 4.1$} & {\color{azure}$4.5 \pm 2.2$} & {\color{azure}$12.0 \pm 2.5$} & {\color{azure}$3.5 \pm 2.1$} & {\color{azure}$17.7 \pm 4.2$} & {\color{azure}$192.1 \pm 18.5$} & -- & --\\
60003 & {\color{azure}$105.2 \pm 26.7$} & {\color{azure}$30.6 \pm 15.2$} & {\color{azure}$<12.5$} & {\color{azure}$27.2 \pm 19.3$} & {\color{azure}$19.0 \pm 12.8$} & {\color{azure}$<12.6$} & {\color{alizarin}$60.4 \pm 30.3$} & {\color{alizarin}$254.9 \pm 83.9$} & {\color{alizarin}$<7.6$} & {\color{alizarin}$123.1 \pm 32.6$}\\
10007 & {\color{azure}$<0.9$} & {\color{azure}$<0.8$} & {\color{azure}$<0.8$} & {\color{azure}$<1.7$} & {\color{azure}$<1.2$} & {\color{alizarin}$<27.3$} & {\color{alizarin}$18.2 \pm 12.6$} & {\color{alizarin}$53.8 \pm 30.6$} & {\color{alizarin}$<9.2$} & {\color{alizarin}$48.4 \pm 9.2$}\\
60000 & {\color{azure}$43.2 \pm 6.4$} & {\color{azure}$12.8 \pm 5.0$} & {\color{azure}$<4.9$} & {\color{azure}$<2.7$} & {\color{azure}$<18.3$} & {\color{azure}$<33.0$} & {\color{alizarin}$24.5 \pm 6.9$} & {\color{alizarin}$190.0 \pm 30.2$} & {\color{alizarin}$<10.6$} & {\color{alizarin}$83.4 \pm 7.7$}\\
60002 & {\color{azure}$10.3 \pm 6.6$} & {\color{azure}$<5.6$} & {\color{azure}$<1.9$} & -- & {\color{azure}$<6.7$} & {\color{azure}$<2.5$} & {\color{alizarin}$<0.5$} & {\color{alizarin}$<0.5$} & {\color{alizarin}$<0.7$} & {\color{alizarin}$<0.6$}\\
60001 & {\color{azure}$133.4 \pm 12.1$} & {\color{azure}$28.1 \pm 6.9$} & {\color{azure}$<5.7$} & {\color{azure}$11.9 \pm 5.4$} & {\color{azure}$27.5 \pm 7.6$} & {\color{alizarin}$<22.1$} & {\color{alizarin}$72.3 \pm 8.1$} & {\color{alizarin}$569.5 \pm 38.6$} & {\color{alizarin}$43.3 \pm 19.3$} & {\color{alizarin}$287.9 \pm 13.1$}\\
80001 & {\color{azure}$<0.2$} & {\color{azure}$<0.3$} & {\color{azure}$<0.6$} & {\color{azure}$<0.5$} & {\color{azure}$<0.5$} & {\color{alizarin}$<0.4$} & {\color{alizarin}$<0.3$} & {\color{alizarin}$<0.3$} & {\color{alizarin}$<0.5$} & {\color{alizarin}$0.8 \pm 0.4$}\\
10001 & {\color{azure}$3.5 \pm 3.0$} & {\color{azure}$3.8 \pm 2.3$} & {\color{azure}$<1.7$} & {\color{azure}$4.8 \pm 2.9$} & {\color{azure}$7.8 \pm 4.0$} & {\color{azure}$<2.6$} & -- & -- & -- & --\\
10010 & -- & -- & {\color{alizarin}$<11.3$} & {\color{alizarin}$2.9 \pm 1.5$} & {\color{alizarin}$6.9 \pm 1.7$} & {\color{alizarin}$2.1 \pm 2.1$} & {\color{alizarin}$13.6 \pm 1.2$} & {\color{alizarin}$115.0 \pm 6.6$} & {\color{alizarin}$<1.5$} & {\color{alizarin}$45.9 \pm 2.6$}\\
10018 & {\color{azure}$<4.1$} & {\color{azure}$<5.9$} & {\color{alizarin}$5.1 \pm 2.5$} & {\color{alizarin}$1.7 \pm 1.4$} & {\color{alizarin}$<3.7$} & {\color{alizarin}$<4.1$} & {\color{alizarin}$7.6 \pm 2.0$} & {\color{alizarin}$44.3 \pm 8.6$} & {\color{alizarin}$<3.4$} & {\color{alizarin}$26.5 \pm 3.3$}\\
2 & {\color{alizarin}$37.2 \pm 4.1$} & {\color{alizarin}$29.9 \pm 3.7$} & {\color{alizarin}$18.0 \pm 3.0$} & {\color{alizarin}$13.2 \pm 1.6$} & {\color{alizarin}$25.1 \pm 2.4$} & {\color{alizarin}$9.9 \pm 1.8$} & {\color{alizarin}$44.4 \pm 4.8$} & {\color{alizarin}$561.1 \pm 23.9$} & -- & --\\
4 & {\color{alizarin}$<2.3$} & {\color{alizarin}$2.5 \pm 1.6$} & {\color{alizarin}$2.9 \pm 2.1$} & {\color{alizarin}$2.2 \pm 1.7$} & {\color{alizarin}$2.9 \pm 1.7$} & {\color{alizarin}$<1.7$} & {\color{alizarin}$6.1 \pm 1.6$} & {\color{alizarin}$33.6 \pm 6.5$} & -- & --\\
3 & {\color{alizarin}$9.2 \pm 2.6$} & {\color{alizarin}$15.3 \pm 2.2$} & {\color{alizarin}$7.7 \pm 2.2$} & {\color{alizarin}$11.0 \pm 3.0$} & {\color{alizarin}$17.2 \pm 2.9$} & {\color{alizarin}$6.1 \pm 2.5$} & {\color{alizarin}$35.1 \pm 4.4$} & {\color{alizarin}$340.6 \pm 22.6$} & -- & --\\
\cutinhead{JADES sample}
18322 & {\color{fuchsia}$<6.5$} & {\color{fuchsia}$10.8 \pm 6.1$} & {\color{fuchsia}$16.5 \pm 12.9$} & {\color{fuchsia}$<5.6$} & {\color{fuchsia}$7.0 \pm 4.4$} & {\color{fuchsia}$4.5 \pm 3.9$} & {\color{azure}$15.6 \pm 4.1$} & {\color{azure}$115.3 \pm 15.2$} & {\color{azure}$<2.7$} & {\color{azure}$47.1 \pm 3.8$}\\
19519 & {\color{fuchsia}$21.1 \pm 5.6$} & {\color{fuchsia}$9.6 \pm 4.5$} & {\color{fuchsia}$11.7 \pm 7.3$} & {\color{fuchsia}$<5.7$} & {\color{azure}$5.7 \pm 3.0$} & {\color{azure}$4.0 \pm 3.6$} & {\color{azure}$16.3 \pm 3.0$} & {\color{azure}$118.5 \pm 11.5$} & {\color{alizarin}$<5.7$} & {\color{alizarin}$48.9 \pm 3.5$}\\
18090 & {\color{azure}$18.0 \pm 7.4$} & {\color{azure}$9.8 \pm 4.3$} & {\color{azure}$<6.7$} & {\color{azure}$<10.3$} & {\color{azure}$9.2 \pm 4.6$} & {\color{azure}$5.1 \pm 4.9$} & {\color{azure}$25.3 \pm 6.2$} & {\color{azure}$214.8 \pm 35.2$} & {\color{alizarin}$<9.6$} & {\color{alizarin}$69.2 \pm 15.1$}\\
22251 & {\color{azure}$11.3 \pm 4.5$} & {\color{azure}$7.5 \pm 3.8$} & {\color{azure}$<4.7$} & {\color{azure}$<7.2$} & {\color{azure}$<6.4$} & {\color{alizarin}$5.6 \pm 4.4$} & {\color{alizarin}$15.5 \pm 3.2$} & {\color{alizarin}$127.0 \pm 13.5$} & {\color{alizarin}$<3.9$} & {\color{alizarin}$52.3 \pm 4.0$}\\
18846 & {\color{azure}$5.7 \pm 3.8$} & {\color{azure}$4.7 \pm 2.8$} & {\color{azure}$<4.8$} & {\color{azure}$4.0 \pm 3.6$} & {\color{alizarin}$4.8 \pm 2.4$} & {\color{alizarin}$2.2 \pm 2.3$} & {\color{alizarin}$14.6 \pm 2.7$} & {\color{alizarin}$109.3 \pm 11.6$} & {\color{alizarin}$<5.5$} & {\color{alizarin}$56.9 \pm 4.5$}\\
10058975 & {\color{alizarin}$3.9 \pm 3.0$} & {\color{alizarin}$4.6 \pm 2.1$} & {\color{alizarin}$2.4 \pm 2.0$} & {\color{alizarin}$<2.0$} & {\color{alizarin}$5.8 \pm 3.0$} & {\color{alizarin}$3.1 \pm 3.0$} & {\color{alizarin}$13.8 \pm 4.1$} & {\color{alizarin}$80.3 \pm 15.9$} & -- & --\\
\enddata
\tablecomments{
Fluxes are in units of $10^{-19}$\,erg/s/cm$^2$. Flux errors and upper limits are $3$-$\sigma$. Lines without the spectroscopic data coverage are marked with ``--". Texts in {\color{fuchsia} purple} are measurements for G140M, {\color{azure} blue} for G235M, and {\color{alizarin} red} for G395M (and G395H for the JADES sample). The presented fluxes are not corrected for magnification nor attenuation. Fluxes for the sources from \citet{nakajima23} are not shown, as we adopt their published values.
}\label{tab:linefluxes}
\end{deluxetable*}

\begin{deluxetable*}{ccccccccccccc}
    \tabletypesize{\footnotesize}
    \tablecolumns{13}
    \tablewidth{0pt}
    \tablecaption{
    Physical properties of the samples
    }
    \tablehead{
    \colhead{ID} & \colhead{R.A.} & \colhead{Decl.} & \colhead{$m_{\rm F444W}$} & \colhead{$z$} & \colhead{$\log$(O/H)} & \colhead{$T_e$} & \colhead{AV$_{\rm neb.}$} & \colhead{$\log M_*$} & \colhead{$\log$\,SFR} & \colhead{$\mu$} & \colhead{$\Delta d_{\rm MSA}$} & \colhead{$\Delta f_{\rm MSA}$}\\
    \colhead{} & \colhead{degree} & \colhead{degree} & \colhead{mag} & \colhead{} & \colhead{+12} & \colhead{$\times10^4$\,K} & \colhead{mag} & \colhead{$M_\odot$} & \colhead{$M_\odot/{\rm yr}$} & \colhead{} & \colhead{kpc} & \colhead{}
    }
    \startdata
    \cutinhead{GTO~1199 Sample}
    30010& $177.387224$ & $22.404619$ & 25.7 & 3.013 & $>7.28$ & $<2.17$ & $0.00$ & $8.81_{-0.02}^{+0.02}$ & $0.7_{-0.1}^{+0.1}$ & $2.1_{-1.6}^{+0.9}$ & $0.50$ & $0.65$ \\
    30055& $177.388125$ & $22.415668$ & 24.9 & 3.214 & $7.91_{-0.14}^{+0.18}$ & $1.58$ & $0.98$ & $8.68_{-0.04}^{+0.04}$ & $0.6_{-0.1}^{+0.1}$ & $5.8_{-1.8}^{+5.9}$ & $0.24$ & $0.66$\\
    20028& $177.410476$ & $22.419310$ & 23.6 & 3.345 & $7.79_{-0.06}^{+0.07}$ & $1.51$ & $0.84$ & $9.79_{-0.04}^{+0.01}$ & $1.8_{-0.1}^{+0.1}$ & $1.4_{-0.1}^{+0.1}$ & $0.57$ & $0.77$\\
    20036& $177.392896$ & $22.412893$ & 28.1 & 3.450 & $>7.15$ & $<3.29$ & -- & $7.67_{-0.05}^{+0.04}$ & $-0.2_{-0.1}^{+0.1}$ & $3.9_{-0.1}^{+0.1}$ & $0.27$ & $0.00$ \\
    150742& $177.429260$ & $22.415850$ & 23.4 & 3.678 & -- & -- & -- & $10.43_{-0.03}^{+0.03}$ & $1.9_{-0.2}^{+0.2}$ & $1.2_{-0.1}^{+0.1}$ & $0.66$ & $0.78$ \\
    20018& $177.399186$ & $22.384109$ & 26.4 & 3.685 & $>6.13$ & $<1.91$ & $0.00$ & $8.56_{-0.05}^{+0.07}$ & $0.1_{-0.1}^{+0.1}$ & $1.9_{-0.1}^{+0.1}$ & $0.36$ & $0.20$ \\
    20014& $177.391600$ & $22.384264$ & 27.2 & 3.693 & -- & -- & -- & $8.26_{-0.12}^{+0.12}$ & $0.2_{-0.1}^{+0.1}$ & $1.6_{-0.1}^{+0.1}$ & $0.49$ & $0.07$ \\
    20024& $177.424202$ & $22.386257$ & -- & 3.696 & $>7.15$ & $<1.95$ & $1.37$ & $10.23_{-0.05}^{+0.06}$ & $1.8_{-0.1}^{+0.1}$ & $2.1_{-0.1}^{+0.1}$ & $0.17$ & -- \\
    20044& $177.383725$ & $22.406203$ & 25.4 & 3.732 & $>7.49$ & $<1.69$ & $0.00$ & $9.11_{-0.03}^{+0.03}$ & $0.7_{-0.1}^{+0.1}$ & $2.9_{-0.1}^{+0.1}$ & $0.54$ & $0.63$ \\
    30093& $177.415100$ & $22.418217$ & -- & 4.194 & $>7.82$ & $<1.70$ & $0.63$ & $9.18_{-0.11}^{+0.04}$ & $1.0_{-0.1}^{+0.1}$ & $1.3_{-0.1}^{+0.1}$ & $0.40$ & -- \\
    150880& $177.403656$ & $22.368355$ & 26.6 & 4.247 & $8.00_{-0.08}^{+0.10}$ & $1.49$ & $0.50$ & $9.28_{-0.04}^{+0.02}$ & $0.9_{-0.1}^{+0.1}$ & $1.3_{-0.1}^{+0.1}$ & $0.29$ & $0.41$\\
    320108& $177.402573$ & $22.384069$ & 23.7 & 4.257 & $8.36_{-0.05}^{+0.05}$ & $1.09$ & $1.16$ & $10.13_{-0.01}^{+0.01}$ & $1.6_{-0.1}^{+0.1}$ & $2.2_{-0.1}^{+0.1}$ & $0.42$ & $0.46$\\
    80002& $177.394550$ & $22.394321$ & 28.4 & 4.495 & $>7.50$ & $<1.69$ & $0.00$ & $7.55_{-0.25}^{+0.34}$ & $-0.6_{-0.1}^{+0.2}$ & $5.7_{-0.4}^{+0.2}$ & $0.19$ & $0.34$ \\
    320002& $177.428253$ & $22.409452$ & 27.5 & 4.658 & $7.41_{-0.16}^{+0.19}$ & $2.59$ & $1.24$ & $8.68_{-0.05}^{+0.08}$ & $0.7_{-0.1}^{+0.1}$ & $1.3_{-0.1}^{+0.1}$ & $0.19$ & $0.39$\\
    320005& $177.422580$ & $22.414459$ & -- & 4.658 & $>8.35$ & $<1.15$ & $1.29$ & $9.60_{-0.02}^{+0.02}$ & $1.4_{-0.1}^{+0.1}$ & $1.3_{-0.1}^{+0.1}$ & $0.55$ & $0.65$ \\
    150903& $177.424210$ & $22.408874$ & 26.4 & 4.659 & $7.90_{-0.09}^{+0.11}$ & $1.68$ & $0.00$ & $9.25_{-0.01}^{+0.01}$ & $0.9_{-0.1}^{+0.1}$ & $1.4_{-0.1}^{+0.1}$ & $0.16$ & $0.34$\\
    60003& $177.375739$ & $22.399017$ & 25.0 & 5.465 & -- & -- & -- & $9.19_{-0.05}^{+0.07}$ & $1.1_{-0.1}^{+0.1}$ & $1.7_{-0.1}^{+0.1}$ & $0.59$ & $0.55$ \\
    10007& $177.420044$ & $22.388666$ & -- & 5.583 & -- & -- & -- & $7.48_{-0.36}^{+0.75}$ & $-0.1_{-0.2}^{+0.2}$ & $3.3_{-0.1}^{+0.1}$ & $0.22$ & -- \\
    60000& $177.378968$ & $22.402419$ & 24.4 & 5.585 & -- & -- & -- & $9.63_{-0.06}^{+0.06}$ & $1.3_{-0.1}^{+0.1}$ & $2.1_{-0.1}^{+0.1}$ & $0.23$ & $0.47$ \\
    60002& $177.382540$ & $22.395407$ & 25.2 & 5.594 & -- & -- & -- & $9.23_{-0.04}^{+0.08}$ & $1.0_{-0.1}^{+0.1}$ & $2.2_{-0.1}^{+0.1}$ & $0.62$ & $0.69$ \\
    60001& $177.412014$ & $22.415776$ & 24.0 & 5.686 & $>7.69$ & $<2.02$ & $1.26$ & $10.37_{-0.02}^{+0.02}$ & $1.7_{-0.1}^{+0.1}$ & $1.5_{-0.1}^{+0.1}$ & $0.58$ & $0.72$ \\
    80001& $177.395860$ & $22.412365$ & 28.8 & 5.974 & -- & -- & -- & $6.35_{-0.73}^{+0.66}$ & $-1.3_{-0.1}^{+0.2}$ & $11.3_{-3.8}^{+12.9}$ & $0.33$ & $0.09$ \\
    10001& $177.388351$ & $22.382391$ & 26.4 & 6.119 & -- & -- & -- & $9.41_{-0.01}^{+0.01}$ & $0.8_{-0.1}^{+0.1}$ & $1.5_{-0.1}^{+0.1}$ & $0.38$ & $0.38$ \\
    10010& $177.412949$ & $22.418858$ & 25.7 & 6.311 & $7.66_{-0.16}^{+0.21}$ & $1.78$ & $0.62$ & $9.59_{-0.03}^{+0.02}$ & $1.0_{-0.1}^{+0.1}$ & $1.4_{-0.1}^{+0.1}$ & $0.17$ & $0.44$\\
    10018& $177.414642$ & $22.417057$ & 27.8 & 6.372 & $>7.28$ & $<2.46$ & -- & $8.73_{-0.07}^{+0.03}$ & $0.4_{-0.1}^{+0.1}$ & $1.4_{-0.1}^{+0.1}$ & $0.41$ & $0.51$ \\
    2& $177.417709$ & $22.417431$ & 23.8 & 7.232 & $7.92_{-0.04}^{+0.04}$ & $1.64$ & $0.00$ & $9.46_{-0.03}^{+0.02}$ & $1.9_{-0.1}^{+0.1}$ & $1.4_{-0.1}^{+0.1}$ & $0.26$ & $0.54$\\
    4& $177.382950$ & $22.412039$ & 27.6 & 7.238 & $>7.21$ & $<2.41$ & $0.00$ & $7.76_{-0.12}^{+0.12}$ & $-0.1_{-0.1}^{+0.1}$ & $3.0_{-0.1}^{+0.2}$ & $0.16$ & $0.03$ \\
    3& $177.389908$ & $22.412710$ & 25.6 & 9.114 & $7.82_{-0.07}^{+0.07}$ & $1.66$ & $0.00$ & $7.77_{-0.02}^{+0.04}$ & $0.5_{-0.1}^{+0.1}$ & $12.5_{-5.0}^{+12.0}$ & $0.31$ & $0.57$\\
    \cutinhead{JADES Sample}
    18090& $53.167180$ & $-27.774615$ & 28.2 & 4.769 & $7.64_{-0.17}^{+0.21}$ & $1.93$ & $0.00$ & $8.58_{-0.11}^{+0.05}$ & $0.4_{-0.5}^{+0.3}$ & -- & $0.37$ & $0.00$ \\
    18322& $53.131707$ & $-27.774119$ & 28.4 & 3.153 & $7.26_{-0.15}^{+0.20}$ & $2.78$ & $0.43$ & $7.29_{-0.16}^{+0.23}$ & $-0.3_{-0.3}^{+0.3}$ & -- & $0.20$ & $0.36$ \\
    18846& $53.134917$ & $-27.772709$ & 27.3 & 6.336 & $7.55_{-0.17}^{+0.22}$ & $1.97$ & $2.70$ & $8.41_{-0.07}^{+0.08}$ & $0.7_{-0.1}^{+0.1}$ & -- & $0.47$ & $0.35$ \\
    19519& $53.166086$ & $-27.771262$ & 27.1 & 3.604 & $7.42_{-0.16}^{+0.21}$ & $2.47$ & $0.39$ & $8.64_{-0.10}^{+0.08}$ & $0.6_{-0.1}^{+0.1}$ & -- & $0.22$ & $0.34$ \\
    22251& $53.154069$ & $-27.766072$ & 27.0 & 5.803 & $7.22_{-0.09}^{+0.18}$ & $3.23$ & $1.39$ & $7.90_{-0.16}^{+0.22}$ & $0.5_{-0.1}^{+0.1}$ & -- & $0.41$ & $0.40$ \\
    10058975& $53.112434$ & $-27.774609$ & 27.5 & 9.435 & $7.11_{-0.15}^{+0.25}$ & $2.98$ & $1.89$ & $8.36_{-0.19}^{+0.32}$ & $0.9_{-0.1}^{+0.3}$ & -- & $0.27$ & $0.00$ \\
    \cutinhead{\citet{nakajima23} Sample}
    ERO\_04590& $110.859818$ & $-73.449127$ & 26.0 & 8.496 & $7.09_{-0.08}^{+0.10}$ & $2.49$ & $1.57$ & $9.05_{-0.21}^{+0.23}$ & $1.1_{-0.1}^{+0.2}$ & $8.7$ & $0.31$ & $0.24$ \\
    ERO\_5144& $110.839674$ & $-73.445357$ & 27.9 & 6.378 & $7.89_{-0.08}^{+0.09}$ & $1.56$ & $0.00$ & $8.75_{-0.15}^{+0.13}$ & $1.2_{-0.7}^{+0.8}$ & $3.2$ & $0.14$ & $0.26$ \\
    ERO\_6355& $110.844917$ & $-73.435043$ & 24.4 & 7.665 & $8.29_{-0.08}^{+0.10}$ & $1.19$ & $0.00$ & $9.07_{-0.12}^{+0.16}$ & $1.5_{-0.1}^{+0.1}$ & $1.8$ & $0.29$ & $0.49$ \\
    ERO\_10612& $110.834351$ & $-73.435059$ & 25.8 & 7.660 & $7.66_{-0.05}^{+0.06}$ & $1.95$ & $0.34$ & $8.54_{-0.16}^{+0.17}$ & $1.0_{-0.1}^{+0.1}$ & $1.9$ & $0.46$ & $0.43$ \\
    GLASS\_100003& $3.604509$ & $-30.380444$ & 25.1 & 7.877 & $7.67_{-0.10}^{+0.15}$ & $2.00$ & $0.00$ & $9.12_{-0.05}^{+0.04}$ & $1.4_{-0.1}^{+0.1}$ & $1.3$ & $0.43$ & $0.40$ \\
    GLASS\_10021& $3.608511$ & $-30.418540$ & 24.7 & 7.286 & $7.88_{-0.08}^{+0.10}$ & $1.68$ & $0.00$ & $8.69_{-0.04}^{+0.15}$ & $1.6_{-0.1}^{+0.1}$ & $1.7$ & $0.47$ & $0.56$ \\
    GLASS\_150029& $3.577166$ & $-30.422576$ & 26.2 & 4.584 & $7.77_{-0.05}^{+0.06}$ & $1.67$ & $0.00$ & $9.31_{-0.03}^{+0.02}$ & $0.9_{-0.1}^{+0.1}$ & $1.5$ & $0.35$ & $0.26$ \\
    GLASS\_160133& $3.580275$ & $-30.424404$ & 26.2 & 4.015 & $8.00_{-0.04}^{+0.05}$ & $1.44$ & $0.05$ & $8.91_{-0.04}^{+0.02}$ & $0.7_{-0.1}^{+0.3}$ & $1.5$ & $0.12$ & $0.40$ \\
    CEERS\_01027& $214.882994$ & $52.840416$ & -- & 7.820 & $7.75_{-0.12}^{+0.17}$ & $1.79$ & $3.18$ & $9.02_{-0.33}^{+0.13}$ & $1.0_{-0.1}^{+0.1}$ & -- & $0.38$ & -- \\
    CEERS\_01536& $214.977227$ & $52.940782$ & -- & 5.034 & $7.64_{-0.09}^{+0.13}$ & $2.15$ & $0.00$ & $8.85_{-1.19}^{+1.19}$ & $1.2_{-0.2}^{+0.2}$ & -- & $0.18$ & -- \\
    \enddata
    \tablecomments{
    $\dagger$: Only those have direct metallicity measurements via \oiiif\ are shown (i.e. Fig.~\ref{fig:linefit}). For those with upper limit on the \oiiif\ flux, we derive a lower limit on \logoh. Errors and upper limits are $1\,\sigma$.
    Electron temperature ($T_e$) in units of $10^4$\,K.
    }\label{tab:phys}
    \end{deluxetable*}

\tabletypesize{\footnotesize} \tabcolsep=0.11cm
\begin{deluxetable}{@{\extracolsep{4pt}}cccccc@{}} 
    \tablecolumns{6}
    \tablewidth{0pt}
    \tablecaption{Mass--Metallicity Relations of Galaxies at $3<z<9.5$. Linear Regression Best-Fit Coefficients}
    \tablehead{
    \colhead{$N$} & \colhead{$\log M_0\,/\,\Msun$} & \colhead{$\alpha$} &
    \colhead{$\beta_z$} & \colhead{$\alpha_z$} & \colhead{$\sigma_\mathrm{\log MZR}$} 
    }
\startdata
\cutinhead{Full Sample}
34 & 8.8  & $0.27_{-0.09}^{+0.09}$ & $7.73_{-0.42}^{+0.41}$ & $0.01_{-0.50}^{+0.50}$ & $0.26_{-0.04}^{+0.06}$\\
\cutinhead{Loss-fraction $<0.5$}
21 & 8.8  & $0.39_{-0.11}^{+0.11}$ & $7.99_{-0.54}^{+0.55}$ & $-0.33_{-0.67}^{+0.66}$ & $0.27_{-0.06}^{+0.07}$
\enddata   
\tablecomments{
Best-fit coefficients for the single slope regression, \logoh\,$=\alpha \log (M_*/M_0) + B(z)$, where $B(z) = \beta_z + \alpha_z \log(1+z)$.
}
\label{tab:param}
\end{deluxetable}

\section*{Acknowledgements}
We wish to thank the anonymous referee for the careful reading of the manuscript and constructive comments. Some/all of the data presented in this paper were obtained from the Mikulski Archive for Space Telescopes (MAST) at the Space Telescope Science Institute. The specific observations analyzed can be accessed via \dataset[10.17909/q8cd-2q22]{https://doi.org/10.17909/q8cd-2q22}. We acknowledge support for this work under NASA grant 80NSSC22K1294. C.G., P.R., and S.S. acknowledge support through grants MIUR2017 WSCC32 and MIUR2020 SKSTHZ.
KB and MT are supported by the Australian Research Council Centre of Excellence for All Sky Astrophysics in 3 Dimensions (ASTRO 3D), through project number CE170100013.

{
{\it Software:} 
Astropy \citep{astropy13,astropy18,astropy22}, bbpn \citep{bbpn}, EAzY \citep{brammer08}, EMCEE \citep{foreman13}, gsf \citep{morishita19}, numpy \citep{numpy}, python-fsps \citep{foreman14}, JWST pipeline \citep{jwst}.
}


\bibliography{sample631}{}
\bibliographystyle{aasjournal}



\end{document}